\documentclass[PAPER, atlasdraft=false,coverpage=false, cernpreprint=true, USenglish, texmf, orcidlogo]{atlasdoc}
\usepackage[subcaption]{atlaspackage}
\usepackage{atlasbiblatex}

\usepackage{atlasphysics}
\graphicspath{{logos/}{figures/}}

\usepackage{ANA-EXOT-2021-19-PAPER-defs}

\AtlasTitle{Search for new physics in final states with semi-visible jets or anomalous signatures using the ATLAS detector}
\AtlasAbstract{%
A search is presented for hadronic signatures of beyond the Standard Model (BSM) physics, with an emphasis on signatures of a strongly-coupled hidden dark sector accessed via resonant production of a $Z'$ mediator.
The ATLAS experiment dataset collected at the Large Hadron Collider from 2015 to 2018 is used, consisting of proton--proton collisions at $\sqrt{s}$ = 13 TeV and corresponding to an integrated luminosity of 140 fb$^{-1}$.
The $Z'$ mediator is considered to decay to two dark quarks, which each hadronize and decay to showers containing both dark and Standard Model particles, producing a topology of interacting and non-interacting particles within a jet known as ``semi-visible".
Machine learning methods are used to select these dark showers and reject the dominant background of mismeasured multijet events, including an anomaly detection approach to preserve broad sensitivity to a variety of BSM topologies.
A resonance search is performed by fitting the transverse mass spectrum based on a functional form background estimation.
No significant excess over the expected background is observed.
Results are presented as limits on the production cross section of semi-visible jet signals, parameterized by the fraction of invisible particles in the decay and the $Z'$ mass, and by quantifying the significance of any generic Gaussian-shaped mass peak in the anomaly region.
}

\AtlasRefCode{EXOT-2021-19}

\PreprintIdNumber{CERN-EP-2025-101}

\AtlasJournal{Phys.\ Rev.\ D.}
\AtlasJournalRef{\PRD 112 (2025) 012021}
\AtlasDOI{https://doi.org/10.1103/44zp-mh1q}

\hypersetup{pdftitle={ATLAS document},pdfauthor={The ATLAS Collaboration}}

\begin{document}

\maketitle

\tableofcontents

% The next lines are included from the .//sections/intro.tex input file
\section{Introduction}
\label{sec:intro}

The particle nature of dark matter (DM) is one of the most pressing questions in fundamental particle physics today.
There is overwhelming experimental evidence from astrophysical experiments for its existence~\cite{rotationcurves, Clowe_2006, 10.1093/mnras/stx3363}, yet no particle in the Standard Model (SM) has all of its observed characteristics.
The range of experimentally allowed DM candidate masses is vast, from axion-like fields at masses as low as $10^{-22}$ eV up to macroscopic galactic objects which can be hundreds of solar masses, and there is a vast experimental landscape attempting to cover these disparate targets~\cite{cirelli2024darkmatter, balazs2024primerdarkmatter}.
Further, DM is just one of several outstanding mysteries in the SM that point to the existence of new physics, motivating a broad search program without strong reliance on signal priors.

General purpose detectors at the Large Hadron Collider (LHC)~\cite{Evans:2008zzb} can provide sensitivity to key areas of beyond the Standard Model (BSM) phase space.
Past collider-based DM searches have primarily focused on weakly interacting massive particles (WIMPs), with masses on the $\mathcal{O}$(100) GeV scale, which can naturally explain the thermal relic density of DM observed today~\cite{Jungman_1996}.
This abundance  calculation depends upon an electroweak interaction between the SM and dark sector, making the search for a mediator particle viable at colliders.
A prominent collider signature is missing transverse energy (\met), due to the non-interacting WIMPs being produced in the proton-proton collision and passing through the detector without interacting.
However, multi-species dark sectors from Hidden Valley theories could yield a more complex picture, predicting a new set of new particles, forces, and interactions with a small coupling to the SM~\cite{Strassler_2007, Han_2008}.

The wide variety of potential dark sector manifestations outlined here are representative of the breadth of phase space that must be considered for a comprehensive search program at colliders.
Anomaly detection (AD) is a strategy that leverages data-driven tools to design search regions that are not designed for a specific signal model, but rather isolate events within a dataset based on their deviation from the learned background distribution~\cite{BELIS2024100091}.
The use of AD-based searches in high energy physics can enable access to potential discoveries not yet predicted by theories, while their inherently generalized design can also allow for more efficient traversal of classes of signal models.

The LHC can provide a unique handle on sensitivity to various BSM scenarios, particularly for strongly interacting dark sectors that produce dark showers and hadronization in LHC detectors, as well as a means for more generic anomaly-based searches.
The theoretical underpinning of focus here is a dark quantum chromodynamics (QCD) with a gauge group $SU(N_d)$ leading to confinement at a scale $\Lambda_d$~\cite{Beauchesne_2019, Bai_2014}.
Dark particles charged under this gauge group would hadronize and generate dark jet signatures.
Some fraction of the dark sector hadrons escape the detector and thus contribute to \met, while others can decay back to Standard Model quarks, which then shower and hadronize according to SM QCD.
Depending on the couplings in this dark sector, a range of jet topologies could be produced, ranging from fully visible or invisible to the detector, and with fully prompt or very displaced tracks as the shower develops.

\begin{figure}[!t]
\centering
\includegraphics[width=0.9\textwidth]{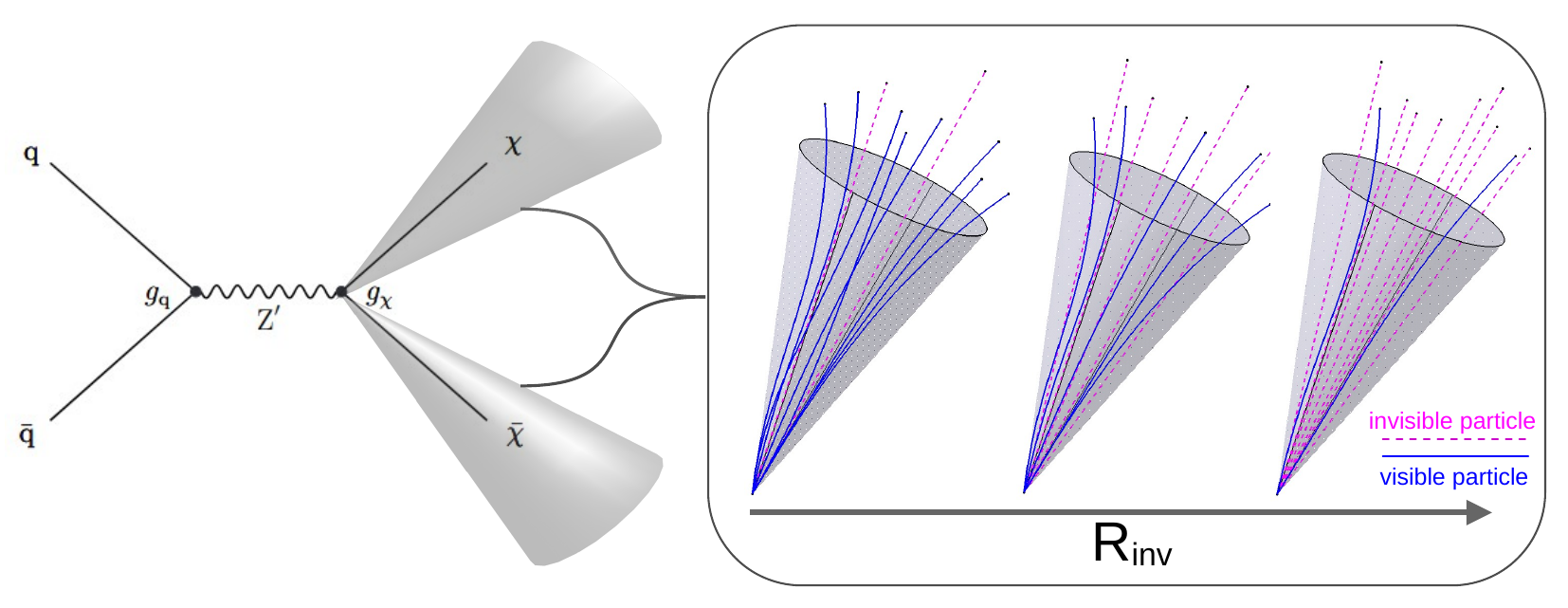}
\caption{Feynman diagram of the signal process, with leading order s-channel production of the $Z'$ mediator which subsequently decays to two dark quarks $\chi$. The dark quarks subsequently generate the semi-visible jet topology upon decay, displaying varying patterns of visible (solid blue) and invisible (dashed pink) constituent energy, coming from SM and dark particles respectively. The relative amount of invisible energy is govern by \rinv, defined as the particle branching ratio to dark hadrons.
\label{fig:diag}}
\end{figure}

This search focuses on a particular signature within this class of models, looking for prompt \textit{semi-visible} jets (SVJs), where a fraction of dark hadrons decay back to SM particles while the others remain invisible, leading to wide jets with high track multiplicity and significant \met~\cite{PhysRevLett.115.171804}.
As the \met~arises from the showering of dark particles in the jet, it will be distributed across the shower, in some cases leading to jet clustering inefficiency.
This class of dark QCD models and their phenomenology have been discussed extensively~\cite{PhysRevLett.115.171804,strassler2008phenomenology,Bernreuther:2019pfb} and signal models with theoretically motivated parameter values have been benchmarked in the 2021 Snowmass community planning process~\cite{snowmass}.
The parameter \rinv~controls the branching ratio of the dark hadron to dark particles, which are non-interacting with the detector, such that higher values of \rinv~lead to larger total missing energy in the jet.
A Feynman diagram of the signal process, along with a graphic depicting the SVJ topology generated upon dark quark decay, can be seen in the diagram in Figure~\ref{fig:diag}.

The complex detector signature of strongly coupled dark sectors motivates the use of machine learning (ML) techniques to isolate them from background processes~\cite{Faucett_2022, Liu_2024}.
In particular, strongly coupled dark sector models present excellent candidates for AD tools due to their sensitive dependence on model parameters~\cite{Barron_2021, Canelli_2022, favaro2025semivisiblejetsenergybasedmodels, pedro2023optimalmassvariablessemivisible, Bardhan_2023, Bai_2024}, with a diverse range of final-state topologies being predicted.
This inherent variability emphasizes the benefit of general and model-agnostic selection strategies to complement more signal model-specific approaches.
Furthermore, the non-perturbative nature of strong dynamics introduces significant challenges in simulation reliability, making data-driven anomaly detection particularly valuable for uncovering potential new physics in these scenarios.

Strongly coupled dark matter searches have recently been a significant topic of interest in collider physics.
The ATLAS experiment published previous searches on fully visible dark jets~\cite{darkQuarks}, a non-resonant production of semi-visible jets~\cite{EXOT-2022-37}, dark mesons decaying to top and bottom quarks~\cite{ATLAS:2024xbu}, and for emerging jets where the dark particles decay back to SM particles with a proper length on the detector scale~\cite{ATLAS_ej}.
The CMS experiment has published searches for resonant production of semi-visible jets~\cite{cms_schan}, as well as for emerging jets~\cite{cms_EJ, CMS:2024gxp}.
As the extreme cases for semi-visible jets involve resonances with fully visible dijet or invisible final states, such searches have potential synergy of exclusions.
Additionally, there has been a broad community effort in the study of anomaly detection methods for model-independent search programs at collider experiments~\cite{Kasieczka_2021, Aarrestad_2022}.
Both CMS~\cite{cmscollaboration2024modelagnosticsearchdijetresonances} and ATLAS~\cite{HDBS-2018-59, HDBS-2019-23, PhysRevLett.132.081801, atlascollaboration2025weaklysupervisedanomalydetection} have performed several anomaly detection searches covering a variety of presentations of BSM physics.

This search focuses on the resonant s-channel production of a $Z'$ mediator particle decaying to two dark quarks which subsequently hadronize, leading to a final state of at least two jets.
While both jets contain dark particles and thus missing energy, only the event-level \met~can be calculated, and any \met~that persists due to imbalance or boosting of the dark jets will be primarily well-aligned with one of them.
The SM background varies with the amount of invisible energy in the SVJs, but is dominated by SM multijet processes.
Discriminating signal from multijet processes is challenging, as the signature of \met~well-aligned with a jet is a feature of jet mismeasurement.
To address these challenges, ML is used to define the signal regions based on high-dimensional and low-level detector information.
In particular, two ML-based signal regions are developed: one based on model-specific classification of SVJ signals from background, and a second leveraging AD to broaden sensitivity to a variety of hadronic signatures with \met.
The final statistical treatment uses a fit to the transverse mass \mt~, calculated using the two leading jets and the \met, in order to reconstruct the $Z'$ mass.
This analysis is unique with respect to previous ATLAS anomaly searches in that it considers new physics signatures that are resonant in \mt, and is the first use of semi-supervised machine learning (with partially labeled training inputs).

% End of text imported from the .//sections/intro.tex input file

% The next lines are included from the .//sections/atlasdetector.tex input file
\section{ATLAS detector}
\label{sec:detector}

\newcommand{\AtlasCoordFootnote}{ ATLAS uses a right-handed coordinate system with its origin at the nominal interaction point (IP) in the centre of the detector and the \(z\)-axis along the beam pipe. The \(x\)-axis points from the IP to the centre of the LHC ring, and the \(y\)-axis points upwards. Polar coordinates \((r,\phi)\) are used in the transverse plane, \(\phi\) being the azimuthal angle around the \(z\)-axis. The pseudorapidity is defined in terms of the polar angle \(\theta\) as \(\eta = -\ln \tan(\theta/2)\) and is equal to the rapidity $ y = \frac{1}{2} \ln \left( \frac{E + p_z}{E - p_z} \right) $ in the relativistic limit. Angular distance is measured in units of \(\Delta R \equiv \sqrt{(\Delta y)^{2} + (\Delta\phi)^{2}}\).}

The ATLAS detector~\cite{PERF-2007-01} at the LHC covers nearly the entire solid angle around the collision point.\footnote{\AtlasCoordFootnote}
It consists of an inner tracking detector surrounded by a thin superconducting solenoid, electromagnetic and hadronic calorimeters,
and a muon spectrometer incorporating three large superconducting air-core toroidal magnets.

The inner-detector system (ID) is immersed in a \qty{2}{\tesla} axial magnetic field
and provides charged-particle tracking in the range \(|\eta| < 2.5\).
The high-granularity silicon pixel detector covers the vertex region and typically provides four measurements per track,
the first hit generally being in the insertable B-layer (IBL) installed before Run~2~\cite{ATLAS-TDR-19,PIX-2018-001}.
It is followed by the SemiConductor Tracker (SCT), which usually provides eight measurements per track.
These silicon detectors are complemented by the transition radiation tracker (TRT),
which enables radially extended track reconstruction up to \(|\eta| = 2.0\).
The TRT also provides electron identification information
based on the fraction of hits (typically 30 in total) above a higher energy-deposit threshold corresponding to transition radiation.

The calorimeter system covers the pseudorapidity range \(|\eta| < 4.9\).
Within the region \(|\eta|< 3.2\), electromagnetic calorimetry is provided by barrel and
endcap high-granularity lead/liquid-argon (LAr) calorimeters,
with an additional thin LAr presampler covering \(|\eta| < 1.8\)
to correct for energy loss in material upstream of the calorimeters.
Hadronic calorimetry is provided by the steel/scintillator-tile calorimeter,
segmented into three barrel structures within \(|\eta| < 1.7\), and two copper/LAr hadronic endcap calorimeters.
The solid angle coverage is completed with forward copper/LAr and tungsten/LAr calorimeter modules
optimised for electromagnetic and hadronic energy measurements respectively.

The muon spectrometer (MS) comprises separate trigger and
high-precision tracking chambers measuring the deflection of muons in a magnetic field generated by the superconducting air-core toroidal magnets.
The field integral of the toroids ranges between \num{2.0} and \qty{6.0}{\tesla\metre}
across most of the detector.
Three layers of precision chambers, each consisting of layers of monitored drift tubes, cover the region \(|\eta| < 2.7\),
complemented by cathode-strip chambers in the forward region, where the background is highest.
The muon trigger system covers the range \(|\eta| < 2.4\) with resistive-plate chambers in the barrel, and thin-gap chambers in the endcap regions.

The luminosity is measured mainly by the LUCID--2~\cite{LUCID2} detector that records Cherenkov light produced in the quartz windows of photomultipliers located close to the beampipe.

Events are selected by the first-level trigger system implemented in custom hardware,
followed by selections made by algorithms implemented in software in the high-level trigger~\cite{TRIG-2016-01}.
The first-level trigger accepts events from the \qty{40}{\MHz} bunch crossings at a rate below \qty{100}{\kHz},
which the high-level trigger further reduces in order to record complete events to disk at about \qty{1}{\kHz}.

A software suite~\cite{SOFT-2022-02} is used in data simulation, in the reconstruction
and analysis of real and simulated data, in detector operations, and in the trigger and data acquisition
systems of the experiment.

% End of text imported from the .//sections/atlasdetector.tex input file

% The next lines are included from the .//sections/datamc.tex input file
\section{Data and simulated samples}
\label{sec:datamc}

This search is performed with 140~\ifb~of LHC $pp$ collision data collected by the ATLAS detector from 2015 to 2018.
The integrated luminosity of the runs is estimated following the methodology described in Ref.~\cite{DAPR-2021-01}. %
It is required that all the relevant elements of the ATLAS detector were fully operational and efficient while the data were collected~\cite{DAPR-2018-01}.

Events are further selected to pass a single-jet trigger selection, where events are required to have a jet at trigger-level with a transverse momentum \pt~exceeding a value that ranges from 360 to 420 GeV depending on the data-taking year.
The lowest \pt unprescaled single-jet trigger is used across data-taking periods.
Selections on jet \pt~(described in Section~\ref{sec:eventsel}) ensure these triggers are only used after reaching an efficiency plateau, avoiding the effect of the trigger turn-on in the analysis.
The data are subject to a blinding strategy throughout the analysis design so as to mitigate analyzer-induced bias, namely that the data of the signal-enriched region are not examined by the analyzers until a full validation of the background estimation has been performed in signal-depleted regions.

Simulated events are generated with a variety of Monte Carlo (MC) generators.
The $pp$ hard scatter physics process is simulated first, and the final-state particles are subsequently showered and decayed.
This full description of the event is then propagated through a detailed detector simulation based on GEANT4~\cite{SOFT-2010-01, Agostinelli:2002hh}, and simulated events are reconstructed with the same algorithms run on collision data.
The generation of the simulated event samples includes the effect of pile-up, defined as the mean number of interactions per bunch crossing.
The effect of pile-up is assessed with the inclusion of overlaid simulated inelastic $pp$ interactions, as well as the effect on the detector response due to interactions from bunch crossings before or after the one containing the hard interaction.
Events in the simulation are weighted by data-taking period in order to reproduce the observed pile-up distribution.

Signal simulation is generated using \textsc{Madgraph5}~\cite{Alwall:2014hca} followed by the Hidden Valley module of \textsc{Pythia8.244}~\cite{Sjostrand:2014zea}.
The s-channel production of $Z'$ is governed by a number of parameters.
The mass of the mediator \mz~can be set, together with the couplings of the $Z'$ to the visible and dark quarks $g_q$ and $g_{\chi}$, respectively.
The dark sector shower is governed by the number of dark colors $N_{c_D}$, the number of dark flavors $N_{f_D}$, and the dark sector confinement scale $\Lambda_D$.
There is also the characteristic scale of the dark hadrons $m_D$ that determines the mass of the dark hadrons, which can be pseudoscalars $\pi_D$ or vectors $\rho_D$, and the mass of the dark quark $m_{\chi}$.
Finally, each dark hadron coming from the dark quark decay or subsequent dark shower has a certain branching ratio to dark (invisible) particles, which is set by \rinv $= \frac{N_{\text{had}_I}}{(N_{\text{had}_I} + N_{\text{had}_V})}$, where $N_{\text{had}_I}$ refers to the number of invisible hadrons and $N_{\text{had}_V}$ refers to the number of visible hadrons.
Signals with higher values of \rinv~thus have larger missing energy in each dark quark originated jet, as well as the event overall.
Up to two additional partons are considered in the underlying process.
The MLM scheme~\cite{Lavesson_2008} for merging tree-level matrix elements and partons showers is used with a $k_T$ merging scale of 30 GeV, a cone radius of 0.4, and max jet $|\eta|$ of 2.5.
The NNPDF2.3LO set of parton distribution functions (PDF)~\cite{Butterworth:2015oua} is used along with the ATLAS A14~\cite{ATL-PHYS-PUB-2014-021} set of tuned parameters (tune).

The values of fixed model parameters are summarized in Table~\ref{tab:model_fixed_params}.
Choices of these values are informed by theoretical considerations and described in detail in Refs.~\cite{Cohen_2017, snowmass}, among others.
The mediator mass \mz~ and the per-dark hadron invisibility probability \rinv~vary, and are used to define the search grid.
The value of \mz~varies between 2000 and 5000 GeV, while~\rinv~ranges from 0.2 to 0.6.

\begin{table}
\caption{Fixed parameters in the semi-visible jet signal model. \label{tab:model_fixed_params}}
\centering
\begin{tabular}{ c|cccccccc }
Parameter & N$_{c_D}$ & $N_{f_D}$ & $\Lambda_D$ [GeV] & $m_{\pi_D}$ [GeV] & $m_{\rho_D}$ [GeV] & $m_{\chi}$ [GeV] & $g_\text{q}$ & $g_\chi$\\
\hline
Value & 3             & 2         & 10        & 17          & 31.77        & 10  & 1 & 0.1 \\
\end{tabular}
\end{table}

The anomaly detection region of the analysis is evaluated and studied with the use of alternate signal models, which do not have SVJ topologies, but share some similar features such as high \met~and resonance in \mt.
One such model is the pair production of new heavy scalars $X$, which each decay to a quark and a dark quark, the latter showering into an emerging jet, where the dark hadrons in the jet have a non-negligible lifetime leading to the presence of displaced tracks within the shower.
The emerging jet signal model consists of five key signal parameters with the following values set: confinement scale $\Lambda_D$ = 1.6 GeV, dark pion mass $m_{\pi_D}$ = $\Lambda_D$/ 2, dark $\rho$ mass $m_{\rho_D}$ = 2$\Lambda_D$ , dark scalar mass $m_{X}$ = 1000 GeV, and decay length of the dark pion $c\tau_{\pi_D}$ = 1 mm.
It is generated with the Hidden Valley module in~\textsc{Pythia8.230} with the NNPDF2.3LO PDF set and the ATLAS A14 tune, and filtered to have at least four jets with \pt~$>$ 100 GeV and $|\eta| <$ 2.7 before detector-level reconstruction.
Additionally, the analysis considers a split-supersymmetric model~\cite{Arvanitaki_2013} with pair production of long-lived gluinos, each decaying to an R-hadron and a lightest supersymmetric particle (LSP) which escapes the detector as \met.
Two points are evaluated from this model, one with a gluino mass of 2000 GeV and another with a gluino mass of 3000 GeV; both points share an LSP mass of 100 GeV and a gluino lifetime of 0.03 ns.
The gluino signals are generated with \textsc{MadGraph 2.6.2}~\cite{Alwall:2014hca}, interfaced to \textsc{Pythia8.24} to simulate decays, parton showering, and the underlying event, using the NNPDF2.3LO PDF set and the A14 tune.

While the background estimation for the analysis is fully data-driven, simulations of multijet SM backgrounds are used to train the ML tools, study modeling of key observables, and develop the statistical treatment.
Multijet processes are generated using
\textsc{MadGraph5} 2.9.9 interfaced to
\textsc{Pythia8.235} for shower and hadronization with the NNPDF2.3LO PDF set and the A14 tune.
The samples are generated in bins of transverse momentum, to ensure good statistics across the momentum spectrum.

% End of text imported from the .//sections/datamc.tex input file

% The next lines are included from the .//sections/objects.tex input file
\section{Object selection}
\label{sec:obj}

Jets are reconstructed from constituent particles via the anti-$k_t$ algorithm~\cite{Cacciari:2008gp} with a radius parameter of 0.4 using the \textsc{FastJet} package~\cite{Fastjet}.
The constituents are particle-flow objects, which are built using a combination of charged particle tracks and calorimeter clusters~\cite{PERF-2014-07, PERF-2015-09}.
Tracks must be matched to the nominal primary vertex~\footnote{The primary vertex is defined as the one with the largest $\Sigma$ \pt$^2$ of its associated tracks.} with the requirement |$z_{0}$ sin${\theta}$| $<$ 2.0 mm. %
The particle-flow construction uses ID track observables to reconstruct the charged particles' four-momentum for \pt$<$100 GeV and calorimeter observables at higher \pt~where the momentum resolution of the tracker deteriorates.

Baseline jets are required to have a minimum \pt~ of 20 GeV and $|\eta| <$ 4.5.
The absolute jet energy scale (JES) calibration is used to correct the jet four-momentum to the particle-level energy scale, as derived using simulated jets in dijet MC events.
Finally, the jet resolution is corrected by applying a Gaussian smearing to simulated jets to match the resolution observed in data~\cite{JETM-2018-05}.

The machine learning methods described in Section~\ref{sec:ml} train over jets using an input modeling of their ghost-associated tracks.
In the ghost-association process, tracks are treated as infinitesimally soft particles by setting their \pt~to 1 eV~\cite{PERF-2012-02}.
Starting from the jet built with the particle-flow algorithm, these infinitesimally soft tracks are clustered in and can then be ghost-associated to a particular jet.
Ghost association is generally superior to geometric matching in dense subjet environments.
For the ML tools, the tracks are required to have \pt $>$ 1 GeV and $|\eta|$ < 2.1.
To ensure good modeling, jets are required to have at least 3 ghost-associated tracks.

While leptons are not used for signal selection, lepton requirements are established for the purposes of vetoing events that contain leptons and for removal of overlapping objects.
Electron candidates are built from the matching of a reconstructed track in the ID with an energy deposit in the electromagnetic calorimeter that has $E_T >$ 7 GeV.
Baseline electrons are required to have a minimum $\pt$ of 20 GeV and $|\eta| <$ 2.47.
Requirements on the transverse and longitudinal impact parameters, defined with respect to the primary vertex, are applied, specifically $|d_0| / \sigma(d_0) <$ 5 (where $\sigma(d_0)$ is the per-track estimated uncertainty on $d_0$) and $|z_0 sin(\theta)| <$ 0.5 mm.
Electrons are also required to pass the $\texttt{Tight}$ identification and $\texttt{Tight\_VarRad}$ isolation criteria~\cite{EGAM-2018-01}.

Muons are reconstructed from combined tracks that incorporate information from both the ID and the muon spectrometer.
The combination involves the fitting of ID hits, calorimeter energy loss, and muon spectrometer tracks.
Baseline muons must have \pt $>$ 20 GeV and $|\eta| <$ 2.5, as well as $|d_0| / \sigma(d_0) <$ 3 and $|z_0 sin(\theta)| <$ 0.5 mm.
Muons are further required to pass the $\texttt{Medium}$ identification and $\texttt{PflowTight\_VarRad}$ isolation criteria~\cite{MUON-2018-03}.
The \met~is computed from the jets, muons, and electrons which pass the baseline requirements.
It comprises a term based on hard objects, specifically the negative vectorial sum of the transverse momenta of all baseline reconstructed objects, added to a soft term constructed from tracks associated with the primary vertex that are not matched to any hard object~\cite{JETM-2020-03}.
After the initial selection, an overlap removal (OR) is applied to deal with the case of a single object being reconstructed as multiple different objects by the detector.
The OR is applied on selected objects before isolation requirements are imposed.
The key overlaps between leptons and jets are resolved in two steps: first, jets within ${\Delta}R$ $<$ 0.2 of electrons are rejected.
Following this, if a reconstructed lepton (electron or muon) is found to be within ${\Delta}R$ $<$ 0.4 of a jet, the lepton is rejected in favor of the jet.

% End of text imported from the .//sections/objects.tex input file

% The next lines are included from the .//sections/machinelearning.tex input file
\section{Machine learning methodology}
\label{sec:ml}

The signal models pursued in this search are particularly well-suited to ML-based selection tools to enrich the signal-to-background ratio in analysis regions.
Two different ML approaches are used in this search, both providing a per-event score that is subsequently used in the event selection to define the relevant analysis regions (see Section~\ref{sec:eventsel} for details).

\subsection{ParticleFlow Network}
\label{subsec:pfn}

The first ML-based region uses a method built to maximize sensitivity to the generated SVJ signal models.
The specific choice of ML model is motivated by two primary considerations: \textit{permutation invariant input modeling} given that collider events are inherently unordered sets of particles, and a \textit{low-level input modeling} (over tracks) to take advantage of the available high-dimensional information to best exploit correlations within the event.

These requirements are satisfied with a Particle Flow Network (PFN)~\cite{pfn}, which is used to model input events as a DeepSet~\cite{zaheer2018deepsets}, in this case an unordered set of tracks.
Given the unordered and variable-length nature of particles in an event, this choice of modeling as a set can enable the ML tool to better learn the most important features of the dataset that enable a signal-to-background classification.
For the specific case of SVJs, track-level modeling further allows the ML model to use the distinctive pattern of particles and \met~within the jet shower, which is a key topological feature to discriminate from multijet background.

Constructing the PFN first involves the learning of a new unordered representation $\Phi$ for each particle.
Permutation invariance is then enforced by summing over the per-particle $\Phi$ representation to create a new permutation invariant, symmetric, and fixed length latent space basis $\mathcal{O}$.
A masking layer is used to suppress any zero-padded inputs, making the architecture length agnostic as well.
Finally, a classifier is trained over this latent space basis to distinguish signal from background, where truth event labels are used to facilitate training.
The trained PFN can then be used to provide an output score for each event that indicates its probability of being from signal or background processes, enabling the definition of a signal-enriched region.

The PFN is trained to separate SVJs from background considering events using a track-level input modeling.
An event-level picture is built by modeling events by the 80 highest-\pt~tracks each of the leading and subleading jet, where the value of 80 is chosen to capture the most significant information of the jet while reducing the input multiplicity with respect to the full track content.
Each track is described using six variables: its four-vector (\pt, $\eta$, $\phi$, E) assuming the track has the pion mass, and the track impact parameters $d_0$ and $z_0$.
These tracks (up to 160 total) are given to the PFN as a single object.
The tracks are re-centered in $\eta$ and $\phi$ such that the average direction of the two leading jets is mapped to $(\eta,\phi) = (0,0)$, aligning the coordinate system with the center of the jet pair.
Each track is scaled to its relative fraction of the total system energy and transverse momentum, enforcing agnosticism to the total energy and transverse momentum of the event.
Finally, each of the six track variables is scaled per-event so that its range is [0,1].

The PFN model used in this analysis consists of two dense layers with 75 nodes each with rectified linear unit \textsc{(ReLU)} activation~\cite{pmlr-v15-glorot11a}, a $\Phi$ latent space dimension of 64, followed by three more dense layers of dimension 75 with \textsc{ReLU} activation, and a final softmax layer~\cite{10.5555/1162264} to determine the event-level classification with categorical cross-entropy loss.
The Adam optimizer is used with an initial learning rate of 10$^{-3}$~\cite{kingma2017adammethodstochasticoptimization}.
The model is developed using 500,000 events at preselection (defined in Section~\ref{sec:eventsel}) of both signal and background divided into a split of train/test/validation corresponding to 78\%:20\%:2\%.
The signal sample consists of an unweighted combination of all SVJ simulated signal points, and the background consists of simulated multijet processes, both described in Section~\ref{sec:datamc}.
Optimization studies were performed to determine the optimal values for best classification performance considering the number of training epochs, batch size, learning rate, number of neurons, and dimensions of the $\Phi$ space; this model represents the optimal choice across these parameters.

After training, the performance of the PFN is evaluated on the test set.
Figure~\ref{fig:mlvars_pfn} shows distributions of the PFN score in data, background multijet simulation, and signal.
The PFN is able to achieve an average area-under-curve of the receiver operating characteristic (ROC) of 0.93 for the combined signal including all simulated points.
This performance was benchmarked against alternate approaches, namely the use of high-level jet variables with a boosted decision tree, which does not perform as well as the PFN and introduces a stronger correlation to energy scale.
As the PFN score shape is not directly used in the analysis, qualitative shape differences between multijet background and data do not impact the analysis.
The crucial metric for agreement between data and simulation is the selection efficiency; here a difference of $<$ 5\% is assessed for the signal region selection of at least 0.6 on the PFN score as defined in Section~\ref{sec:eventsel}.
A diagram of the PFN can be seen in the upper half of Figure~\ref{fig:mlarch}.

\begin{figure}[!t]
\centering
\includegraphics[width=0.55\textwidth]{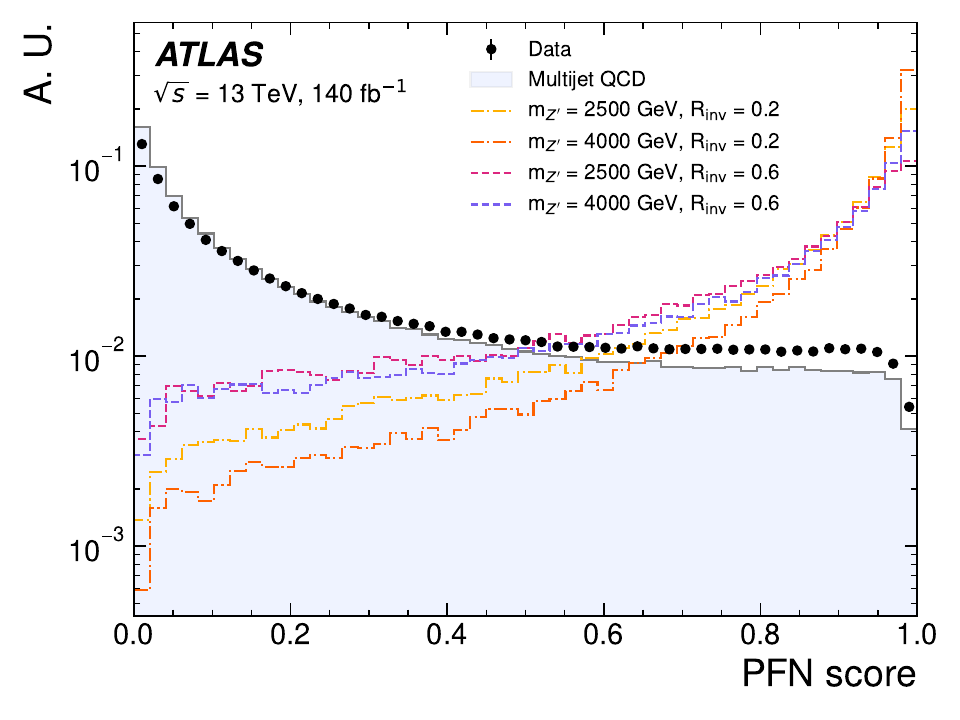}
\caption{Distributions of the PFN score in data, multijet simulation, and representative SVJ signals, after application of the analysis preselection (defined in Section~\ref{sec:eventsel}). All histograms are normalized to unity. The PFN provides good discrimination between signal and background.
\label{fig:mlvars_pfn}}
\end{figure}

\subsection{Semi-supervised anomaly detection}
\label{subsec:antelope}

To broaden the coverage of the analysis, an anomaly detection tool is developed as a companion to the model-focused PFN tool described above.
Anomaly detection involves the application of algorithms that identify abnormalities within a dataset based on their incompatibility with learned Standard Model properties, offering a data-driven way to pursue the unknown unknowns in the BSM landscape and thus expand discovery reach at high energy physics experiments.
\begin{figure}[!t]
\centering
\includegraphics[width=0.95\textwidth]{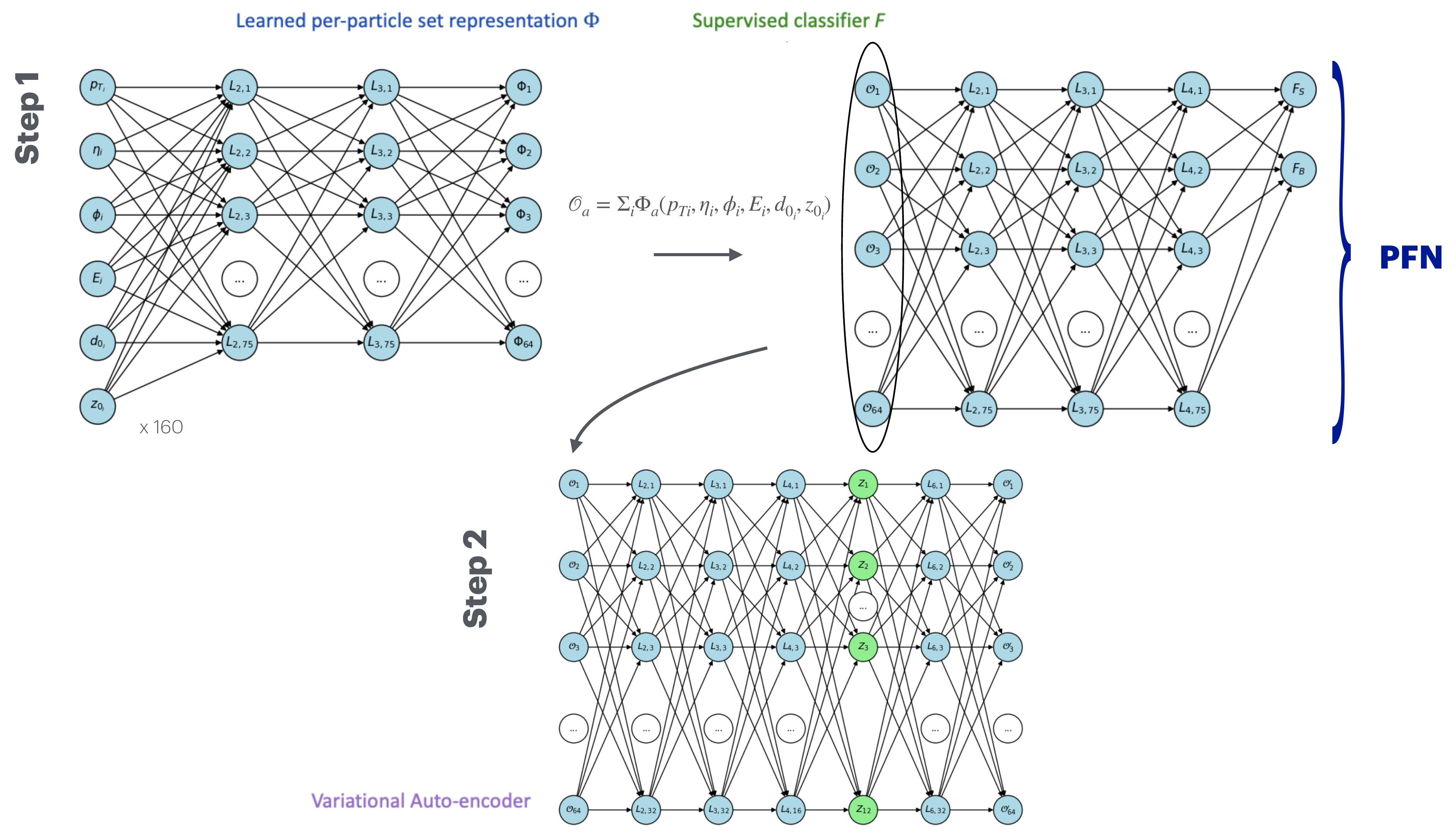}
\caption{An annotated diagram of the ANTELOPE architecture~\cite{matos2024semisupervisedpermutationinvariantparticlelevel}. Step 1 refers to the fully supervised training of the PFN network to classify semi-visible jet signals from SM multijet background. In Step 2, data events are encoded in the permutation-invariant basis $\mathcal{O}$ obtained from the PFN training, and used for unsupervised training of a VAE, where Z nodes specify the latent space.
\label{fig:mlarch}}
\end{figure}

A typical component of an AD analysis is the use of data-driven classification tools, which mitigates reliance on simulation.
However, this requires an approach to training the ML model that does not require full and true labeling of training inputs.
The training metholodogy here is a hybrid approach between full labeling (supervised) and no labeling (unsupervised) known as semi-supervised, where only some training events have correct labels.
This approach informs the AD tool with partial knowledge of the signal model, honing it to the specific task of detecting dark jets within the broad task of classifying anomalous events.
Additionally, the key event characteristics discussed above, namely the relevance of modeling events as unordered and variable-length sets of tracks, motivate a similar input modeling for the AD tool.

With these considerations in mind, a new tool is developed for this analysis, referred to as ANomaly deTEction on particle fLOw latent sPacE (ANTELOPE)~\cite{matos2024semisupervisedpermutationinvariantparticlelevel}, to perform particle-level and permutation-invariant AD on collider events.
In this way, the ANTELOPE model is trained in two stages.
The first is the fully supervised PFN that was already trained to classify the combined SVJ signal from simulated multijet background.
The pre-trained PFN is subsequently used to encode data events into the basis $\mathcal{O}$, and in this format they are given as input to the training of a variational autoencoder (VAE)~\cite{kingma2022autoencodingvariationalbayes}.
A VAE is a common tool that can perform AD via fully unsupervised training by learning to reconstruct data events after forcing them through a lossy compression that motivates the model to extract the most salient features of the inputs, thus learning the underlying data distribution.
The ANTELOPE technique leverages permutation invariance and a signal model to learn a new representation for the events.
Embedding each data event into this representation for VAE training thus leads to a more performant AD method, as compared to a VAE using an input modeling based on fixed-length track features.

Figure~\ref{fig:mlarch} provides a diagram of the ANTELOPE architecture.
The specific ANTELOPE model used here takes the same inputs as the PFN described in the previous section, namely the 80 highest \pt~tracks from each of the leading jets, each described by six variables.
The VAE model used for the ANTELOPE architecture is trained (tested) with 80\% (20\%) of a subset of 200,000 data events subject to preselection requirements (detailed in Section~\ref{sec:eventsel}).
The loss $\mathcal{L}$ is that of the standard VAE, namely the sum of two terms: the mean-squared error (MSE) of the output and the input to test reconstruction quality, and the Kullback-Leibler divergence (KLD) measuring the distance between the learned and tested distributions in the latent space.
The VAE is trained for 50 epochs with a learning rate of 10$^{-5}$ on data events only.
The final ANTELOPE score is produced by applying a log followed by sigmoid transformation function to the total evaluated $\mathcal{L}$.
Figure~\ref{fig:mlvars_ant} shows the distribution of the ANTELOPE score in data and signal simulation with all preselection requirements imposed.

\begin{figure}[!t]
\centering
\includegraphics[width=0.55\textwidth]{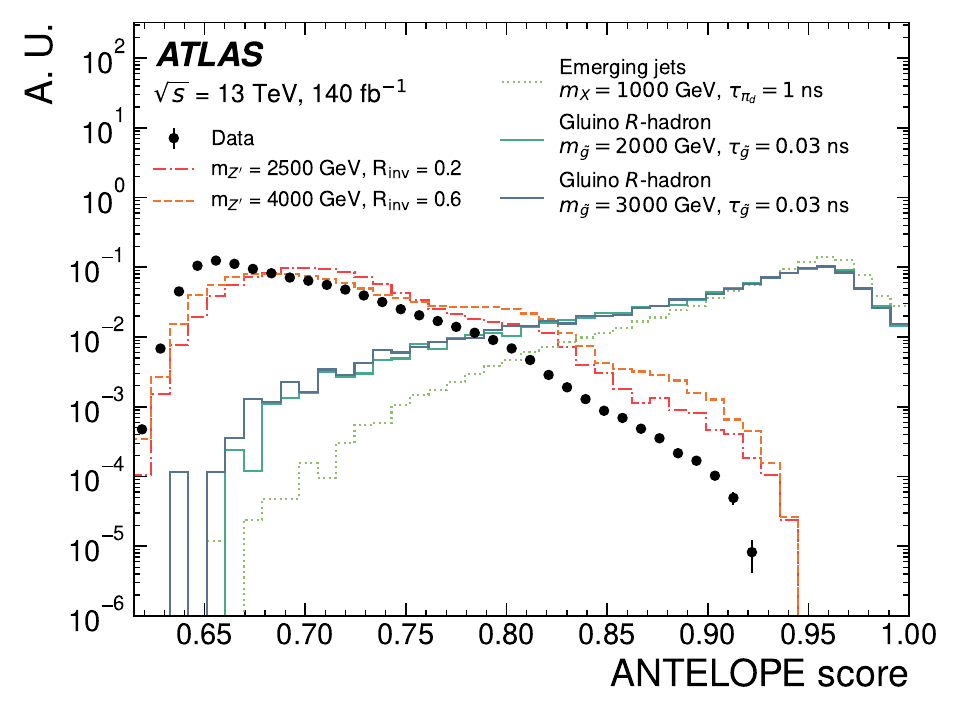}
\caption{Distribution of the ANTELOPE score in data, SVJ signals, and alternate signal models at preselection, where higher values of the ANTELOPE score indicate higher likelihood of anomaly. All histograms are normalized to unity. Though the ANTELOPE training never sees emerging jet or gluino R-hadron signals in its training, it is still able to distinguish them as more anomalous than the data.
\label{fig:mlvars_ant}}
\end{figure}

The performance of the ANTELOPE tool is evaluated through its sensitivity breadth to a wide variety of signal models rather than its depth to the SVJ signal in particular.
To quantify this, several additional simulated signals are considered as test sets, and a selection of $>$ 0.7 is imposed on the ANTELOPE score to maximally enrich the signal sensitivity for the SVJ simulated samples (defined in Section~\ref{sec:eventsel}).
This threshold for the ANTELOPE SR, determined with no awareness of the alternate signal models, is found to enrich the signal sensitivity with respect to the PFN SR cut by over three times for the emerging jet signal, and to the gluino R-hadron signals by greater than a factor of five.
While it is inherently difficult to interpret the outcomes of a less-than-supervised ML tool, studies of the ANTELOPE score along with traditional observables indicate that it is correlated to track displacement variables, as well as event-level kinematic quantities such as missing energy and \mt.

% End of text imported from the .//sections/machinelearning.tex input file

% The next lines are included from the .//sections/eventselection.tex input file
\section{Event selection}
\label{sec:eventsel}

Events are required to have at least one $pp$ interaction vertex with at least two tracks with \pt~greater than 500 MeV~\cite{ATL-PHYS-PUB-2015-026}.
An event-level jet cleaning is applied to reduce the presence of beam-induced background, cosmic rays, and calorimeter noise bursts~\cite{ATLAS-CONF-2015-029}.
A tight working point explicitly designed to provide higher background rejection is used given the similarity of the semi-visible jet signal to non-collision backgrounds.
In addition to the standard cleaning, events containing jets in the bad hadronic calorimeter tiles have been removed from the data sample~\cite{DAPR-2018-01}.
A common preselection is applied to all events considered in the analysis.
In order to reconstruct the resonance mass, events must have at least 2 jets, each containing at least 3 tracks to ensure good modeling.
Since the signal process for this analysis does not directly produce leptons, events containing leptons are vetoed from the data sample.
To ensure full efficiency of the jet trigger, a requirement on the \pt~of the leading jet of at least 450 GeV is imposed.
The subleading jet is required to have \pt $>$ 150 GeV; in conjunction with a selection that the $\Delta\phi$ between the leading two jets be greater than 0.8, this is found to suppress contributions from non-collision background.
Both leading jets must have $|\eta| <$ 2.1, to ensure full tracking coverage.
The absolute value of the difference in rapidity between the two leading jets, $|\Delta$y$|$, must be $<$ 2.8 to ensure final state objects that are central in $\eta$, an event characteristic that is associated to heavy resonance production.
A selection of \met $>$ 200 GeV is imposed to focus on events with the key semi-visible jet characteristic of high missing energy due to dark particles in the shower.
Finally, \mt~is required to be greater than 1.5 TeV, to ensure that signal fits are performed on a smoothly falling background.

A signal region (SR) is defined to enrich the presence of signal and remove SM background processes.
The background estimation in the SR is constructed using a control region (CR), which is defined to provide an orthogonal set of data events enriched in background and similar in kinematic phase space to the SR.
A validation region (VR) is further defined to provide additional validation of the background estimation strategy.
A diagram summarizing the analysis regions and demonstrating the analysis flow can be seen in Figure~\ref{fig:analysisflow}, with details of the selections provided below.
\begin{figure}[!t]
\centering
\includegraphics[width=1.0\textwidth]{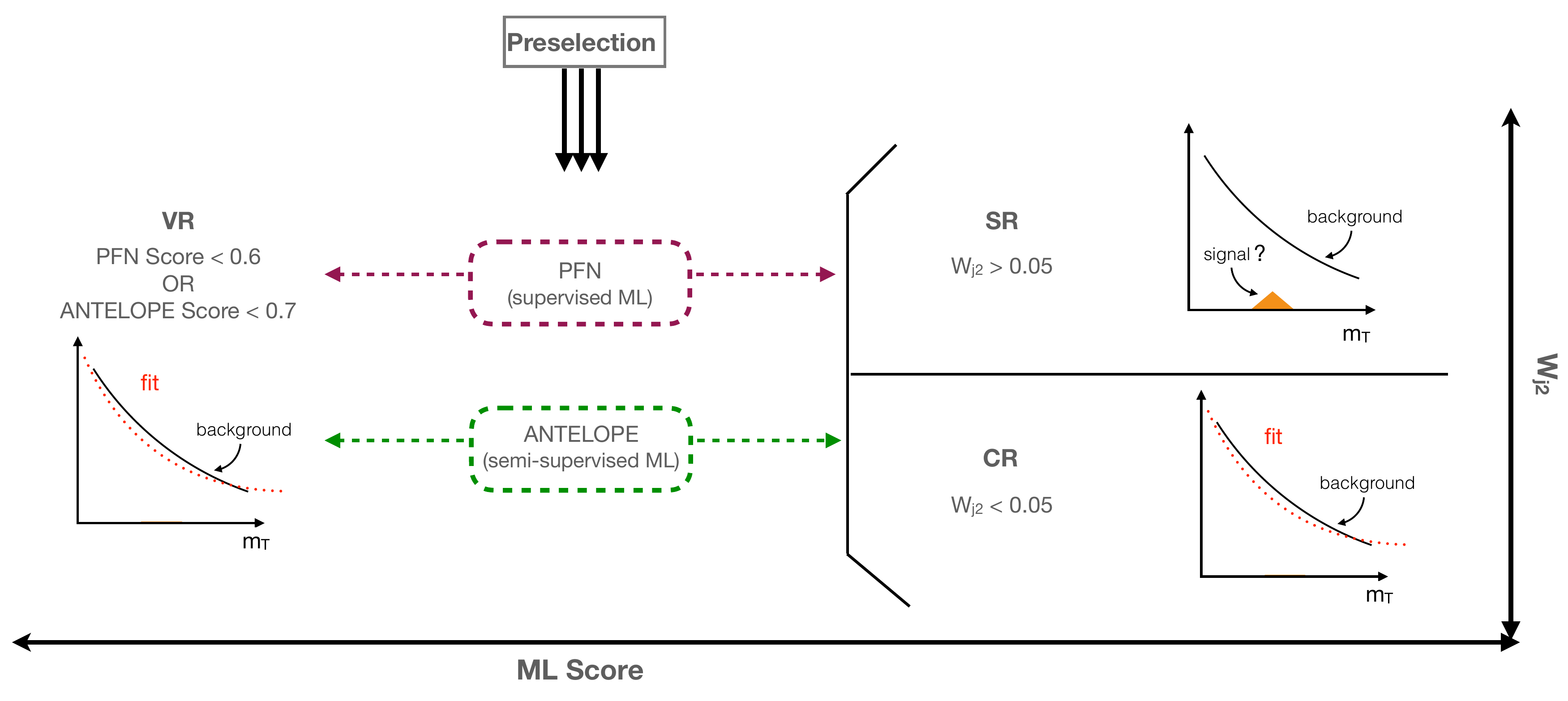}
\caption{Diagram indicating the analysis selections and background estimation/validation fitting strategy, where \mt~refers to the transverse mass calculated with the leading two jets and \met, \width~refers to the width of the subleading jet, and PFN (ANTELOPE) score refer to the evaluated output of the PFN (ANTELOPE) models. The CR and VR fit validations as well as the unblinded SR fit are done twice, once for each of the ML regions.
\label{fig:analysisflow}}
\end{figure}

Three sensitive analysis variables are used to define the various analysis regions.
Dark showers tend to have higher track multiplicty and broader showers than SM multijet processes, resulting in good signal-background discrimination from the jet width, defined by the distance between the calorimeter clusters and the jet axis scaled by the cluster energy:
\begin{equation*}
\label{eq:width}
W_{\text{j}} = \frac{\Sigma_i p_\text{T}^i \times \Delta R_{i,\text{jet}}}{\Sigma_i p_\text{T}^i} %
\end{equation*}
where $i$ refers to a sum over calorimeter clusters and $\Delta R_{i,\text{jet}}$ is the distance in $\eta$-$\phi$ space between calorimeter cluster $i$ and the jet axis~\cite{PERF-2011-03}.

In particular, the subleading jet width \width was found to provide discriminating power for all signals across \rinv~values, allowing for the definition of a signal-enriched region and a corresponding CR with a less than 1\% signal contamination across all simulated signal grid points.
A requirement of \width~$>$ 0.05 is used to define the SR, with the orthogonal requirement of \width~$<$ 0.05 defining the CR.

The PFN and ANTELOPE scores (collectively referred to as ``ML scores") as defined in Section~\ref{sec:ml} are also highly useful to isolate signal-like events.
Selections on both ML scores are chosen to maximize signal sensitivity across the (\mz, \rinv) grid, leading to SR requirements of PFN score (ANTELOPE score) to be greater than 0.6 (0.7) to define the \srpfn~ (\srad).

Figure~\ref{fig:analysisvars} shows distributions of \width, \mt, and \met~in data, as well as in background and signal simulation.
A fundamental assumption of the background estimation strategy is that \width and ML scores are not strongly correlated to \mt, to ensure no substantial sculpting of \mt~could arise in the SR that would weaken the fit studies performed in the CR and VR; this is observed to be true in data at preselection.
A summary of region selections can be found in Table~\ref{tab:selections}.

\begin{figure}[!t]
\centering
\begin{subfigure}[b]{0.45\textwidth}
\centering
\includegraphics[width=\textwidth]{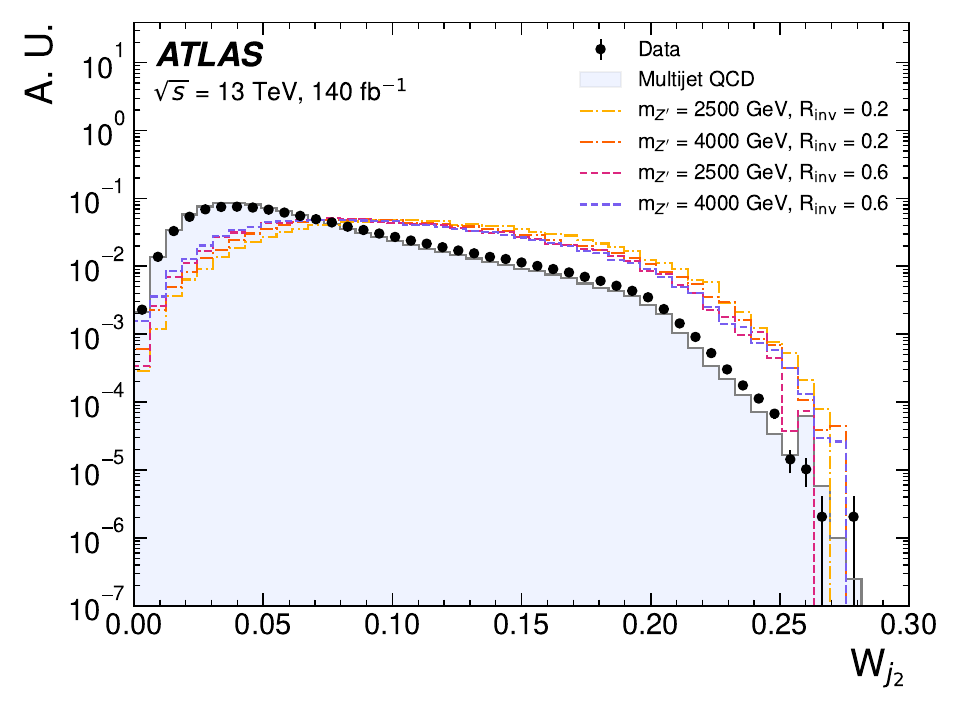}
\label{fig:width_distribution}
\caption{}
\end{subfigure}
\hfill
\begin{subfigure}[b]{0.45\textwidth}
\centering
\includegraphics[width=\textwidth]{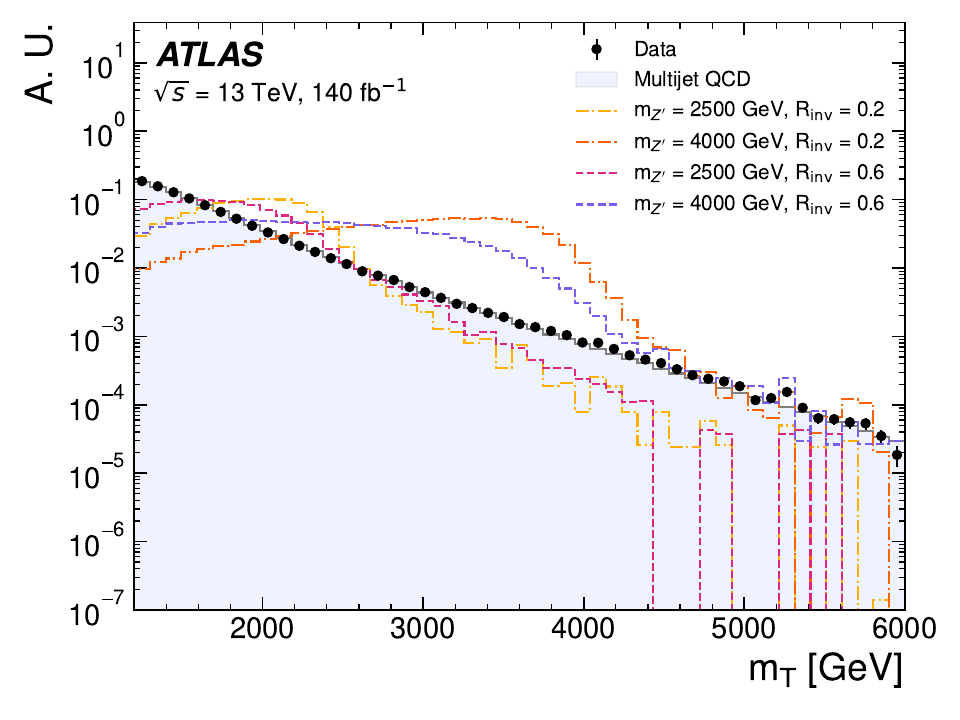}
\label{fig:mt_distribution}
\caption{}
\end{subfigure}
\hfill
\begin{subfigure}[b]{0.45\textwidth}
\centering
\includegraphics[width=\textwidth]{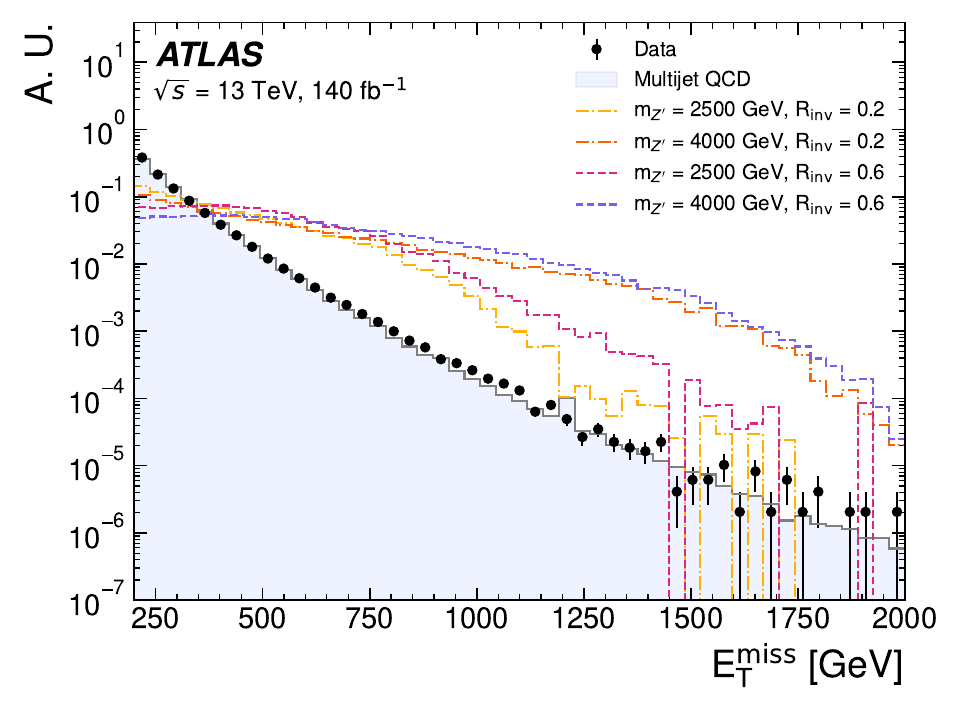}
\label{fig:met_distribution}
\caption{}
\end{subfigure}
\caption{Distributions of analysis variables (a) \width, (b) \mt, and (c) \met in data, background simulation, and signal at preselection. All histograms are normalized to unity. }
\label{fig:analysisvars}
\end{figure}

\begin{table*}[t]
\caption{Preselection requirements, as well as optimized requirements defining the SRs and background estimation regions. \label{tab:selections}}
\centering
\begin{tabular}{|c|c|c|c|c|c|}
\hline
Variable & \multicolumn{5}{c|}{Preselection requirements} \\ \hline
N$_{\text{jets}}$ & \multicolumn{5}{c|}{ $\geq$ 2 }  \\
N$_{\text{tracks}}$ (jet)  & \multicolumn{5}{c|}{ $\geq$ 3 }  \\
N$_{\text{lep}}$ & \multicolumn{5}{c|}{ = 0 }  \\
$p_\text{T} j_1 (j_2)$ \ [GeV] & \multicolumn{5}{c|}{ $>$ 450 ($>$ 150)} \\
$\Delta \phi$ ($j_1$, $j_2$) & \multicolumn{5}{c|}{$>$ 0.8} \\
$|\eta_{j_1,j_2}|$ & \multicolumn{5}{c|}{$<$ 2.1} \\
$\Delta$ y & \multicolumn{5}{c|}{$<$ 2.8} \\
\met[GeV] & \multicolumn{5}{c|}{$>$ 200} \\
\mt[GeV] & \multicolumn{5}{c|}{$>$ 1500} \\
\hline \hline
& \srpfn & \srad &  VR$_{\text{PFN}}$ & VR$_{\text{AD}}$ & CR \\
\hline
\width & \multicolumn{4}{c|}{$>$ 0.05} & $<$ 0.05  \\
\hline
PFN score & $>$ 0.6 & -- & $<$ 0.6 & -- & --  \\
\hline
ANTELOPE score & -- & $>$ 0.7 & -- &$<$ 0.7 & --  \\
\hline
\end{tabular}
\end{table*}

% End of text imported from the .//sections/eventselection.tex input file

% The next lines are included from the .//sections/background.tex input file
\section{Background estimation}
\label{sec:bkg}

The SM background in the SR is predominantly composed of multijet events, and due to the poor modeling of QCD by MC, it is estimated in a fully data-driven way.
An empirical functional form is used for the background shape of \mt.
The ability of this function to model the background behavior is tested both in MC and in data (CR and VR) and the shape parameters are left free in all the fits.
Similarly, for the final results, the shape parameters will be determined from the data fit in the SR.

The fits are performed for \mt~$>$ 1500 GeV.
The functional form is chosen for its fitting performance among other lower/higher order functions and was used in several dijet search analyses~\cite{EXOT-2018-32, Aaboud:2312376}:
\begin{equation*}
f(x) = p_1(1-x)^{p_2}x^{p_3+p_4 \text{ln}x+p_5\text{ln}^2x}
\end{equation*}
where x = m$_{T}$/$\sqrt{s}$ and the $p_i$ are free parameters.
For \srpfn, the \mt~distribution is fit to 90 even-width bins, while \srad~uses 180 bins to ensure good granularity for the subsequent rebinning incorporating the signal mass resolution that is described in Section~\ref{sec:stats}.

% End of text imported from the .//sections/background.tex input file

% The next lines are included from the .//sections/systematics.tex input file
\section{Systematic uncertainties}
\label{sec:syst}

As the background estimation procedure relies on fitting a functional form to the data, any smooth variation on the \mt~spectrum itself is absorbed by the background fit and the impact on the residuals is expected to be small.
Any remaining sources of systematic uncertainty apply to either the signal shape or normalization.

A background uncertainty is ascribed to the possible fitting of spurious signal from fluctuations in the \mt~spectrum.
This systematic is assessed as a normalization uncertainty on the signal model shape by performing a signal-plus-background fit on blinded data events from the signal-depleted CR.
Beforehand the CR data is subject to a smoothing procedure based on functional decomposition to accommodate statistical fluctuations at high \mt~\cite{edgar2018functional}.
The amount of fitted spurious signal varies with \rinv~ and \mz, but is consistently below a level of 0.5$\sigma$, in line with recommendations~\cite{ATL-PHYS-PUB-2020-028}.
This value is added as a number of events that can be attributed to the spurious signal, which is custom for each simulated signal point.

A second normalization uncertainty is ascribed to the impact of theoretical modeling effects, of which three sources are considered.
The first arises from variations to the strong coupling constant $\alpha_s$ and parton distribution functions (PDFs), and the corresponding uncertainty is estimated through study of the signal \mt~distribution under alternate PDF sets and taking an envelope of these variations in the \mt~window corresponding to the signal $Z'$ mass, as prescribed by the PDF4LHC group~\cite{Butterworth:2015oua}.
Second, effects of generator-level variations of the A14 tune on the initial-state and final-state radiation as well as multiple parton interactions, are estimated by varying the strong coupling constant (\alphas), the renormalization scale (\muR), and the factorization scale (\muF) and similarly evaluating the envelope in the relevant \mt~window for a given signal.
Both of these theoretical sources of uncertainty are modeled as a flat 20\% variation on the theoretical cross section of the signal.
The uncertainty in the combined 2015–2018 integrated luminosity is 0.83\%, obtained using the LUCID-2 detector~\cite{LUCID2} for the primary luminosity measurements, complemented by measurements using the ID and calorimeters.

Systematic uncertainties that affect the shape of the signal in \mt~come from instrumental effects, namely in the small-R jet \pt~scale.
Such uncertainties are potentially impactful on an analysis that looks for localized excesses on top of a smoothly falling background because they can shift the peak of the resonance.
Jet \pt~scale uncertainties are estimated using a methodology that compares track-to-calorimeter \pt~double ratios from data to simulation, from which baseline differences are ascribed to systematic effects~\cite{JETM-2018-05}.
In this analysis, the total jet \pt~scale uncertainty has a very small effect on the shape of the \mt~spectrum, but variations are nonetheless incorporated into the final signal-plus-background fits as nuisance parameters with Gaussian constraints.

% End of text imported from the .//sections/systematics.tex input file

% The next lines are included from the .//sections/stats.tex input file
\section{Statistical analysis}
\label{sec:stats}

The statistical treatment of both the \srpfn~and \srad~are developed through functional fitting of the \mt~spectrum.
Studies of the signal mass peak in \mt~using double-sided Crystal Ball functions reveal that the mass resolution varies from approximately 200 GeV for \mz~=2000 GeV and up to 800--1000 GeV for higher \mz, depending on the \rinv~value.
Each \mt~fit is performed with the five functional parameters of the background function floating, including the background normalization.
The CR and VR are used to validate and test the function's capability to fit different \mt~spectra, but the final \srpfn~and \srad~functions are obtained from a new fit of each corresponding SR \mt~distribution.
Systematics are modeled as nuisance parameters with Gaussian constraints, and can be either yield or shape uncertainties, as discussed in Section~\ref{sec:syst}.

A variety of fit tests are performed with the CR and VR data samples to ensure good quality of the background expectation.
The functional form in Section~\ref{sec:bkg} is used to perform a background-only fit on both simulated background and data in the CR and VR selection.
To ensure that the fit tests are done on a region with similar expected statistics to that of the SR, the SR yield is unblinded and the fit test region is randomly downsampled to match the SR event yield.
As the overall event count in the SR does not incorporate the sensitive \mt~shape information, this can be considered to not fully unblind the analysis during fit development and risk the introduction of bias.

To further validate the fit in more instances, pseudo-data is created from the CR/VR data distributions.
It is created following the Asimov prescription with smoothing to accommodate fluctuations in the high \mt~tail.
The smoothing applied follows the procedure for functional decomposition described in Ref.~\cite{edgar2018functional}.
Toys thrown from the resulting smoothed template produce distributions of background-only p-values that are consistent with good modeling.
A final fit check is performed by injecting signal into the CR data template.
These injections range in significance in the $Z'$ mass window from 1$\sigma$ to 5$\sigma$, and a good linearity is observed of the post-fit extracted number of signal events with respect to the injected number.

\paragraph{\srpfn}
Signal interpretations are extracted from \srpfn~ only via signal-plus-background fits using a histogram of the SVJ signal template.
The parameter of interest in the statistical analysis is the signal strength $\mu$, defined as a scale factor multiplying the nominal yield predicted by a 1~pb signal cross section so as to match the observed number of signal events.
The background-only hypothesis corresponds to $\mu$ = 0.
To accurately perform the fit, equidistant narrow-width bins are required as the fit function is performed with \textsc{RooPDF} and evaluates the function at the center of the bin rather than integrating across the bin~\cite{ATL-PHYS-PUB-2020-028}.
Therefore, a narrow binning (90 across the \mt~spectrum) is chosen to perform the functional form background-only and signal-plus-background \mt~fits for \srpfn.

\paragraph{\srad}
As \srad~is built using an anomaly detection tool, it is not used to perform any signal-plus-background fits.
However, the generality of the tool used in its construction results in a region with broad expected sensitivity to new physics producing resonances in \mt.
Therefore, \bh~\cite{choudalakis2011hypothesistestingtrialsfactor} is used to quantify the significance of adjacent bins in \mt~without a particular signal model shape.
\bh~looks for significant excesses in the \mt~spectrum incorporating only the statistical uncertainty of the data and the adjacency of bins that disagree between data and expectation.
It outputs a p-value that provides a goodness-of-fit metric to quantify the most significant excess across the spectrum, along with a window in \mt~corresponding to this excess.
To do this, the 180 narrow bins used for the functional fit are subsequently rebinned to a wider variable binning that reflects the expected signal mass resolution, informed by the double-sided Crystal Ball mass peak fits, and thus increases in width for higher \mt.
There is potential for a spurious signal in the \srad~\mt~spectrum that can be found by the \bh~procedure.
To evaluate this, \bh~fits are performed over Asimov data in the ANTELOPE CR and VR.
In 100 trials, no fits are found with \bh~$p$-values less than 0.01, indicating a small impact of potential spurious signal effects.

% End of text imported from the .//sections/stats.tex input file

% The next lines are included from the .//sections/results.tex input file
%

The unblinded \mt~distributions for the PFN and ANTELOPE signal regions can be seen in Figure~\ref{fig:unblinded_mt}.
The fit quality of the background function to the data is good, with $p$-values of 0.26 and 0.74 for the \srpfn~and \srad~ respectively.
No significant excesses of the data with respect to the background-only model are observed.
\bh~is run on the \srad~\mt~distribution rebinned according to expected signal mass resolution as described in Section~\ref{sec:stats} allowing for excesses with widths of at least 2 and at most 6 bins.
The \bh~interval of the \srad~\mt~spectrum with the most significant excess is found to be between 1700 and 1900 GeV, with a corresponding \bh~p-value of 0.81, indicating that no significant Gaussian resonance shape can be found in the data.

\begin{figure}[!t]
\centering
\begin{subfigure}[b]{0.45\textwidth}
\centering
\includegraphics[width=\textwidth]{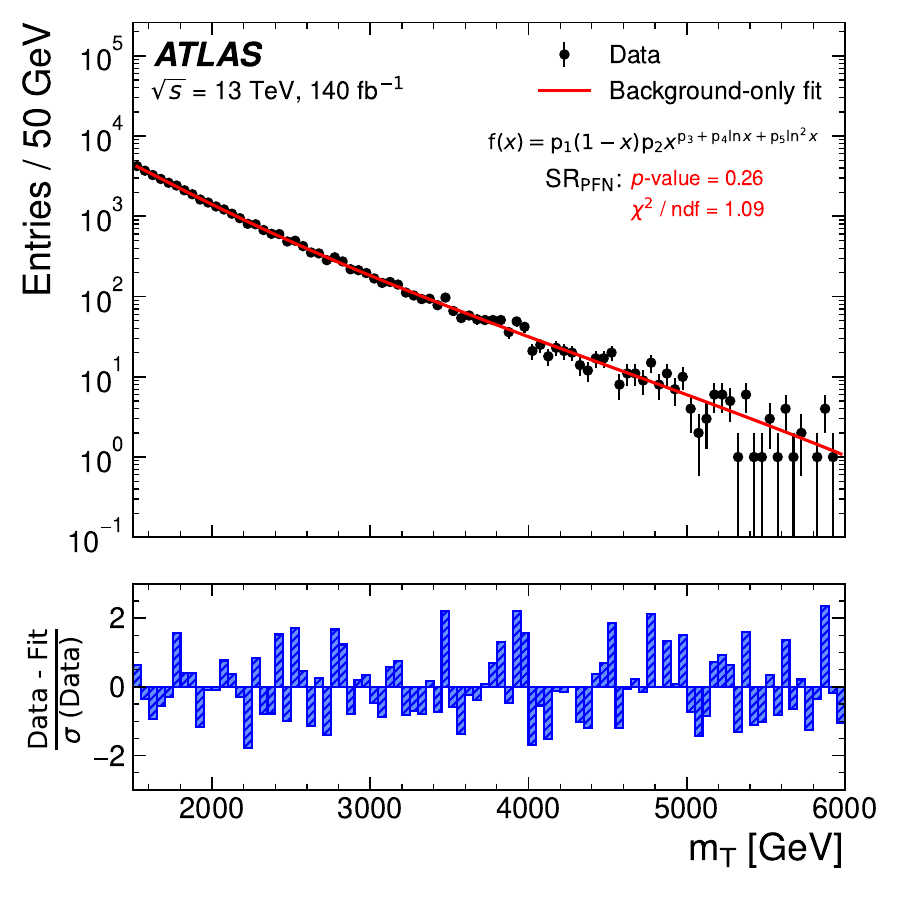}
\caption{}
\label{fig:bkg_fit_PFN}
\end{subfigure}
\hfill
\begin{subfigure}[b]{0.45\textwidth}
\centering
\includegraphics[width=\textwidth]{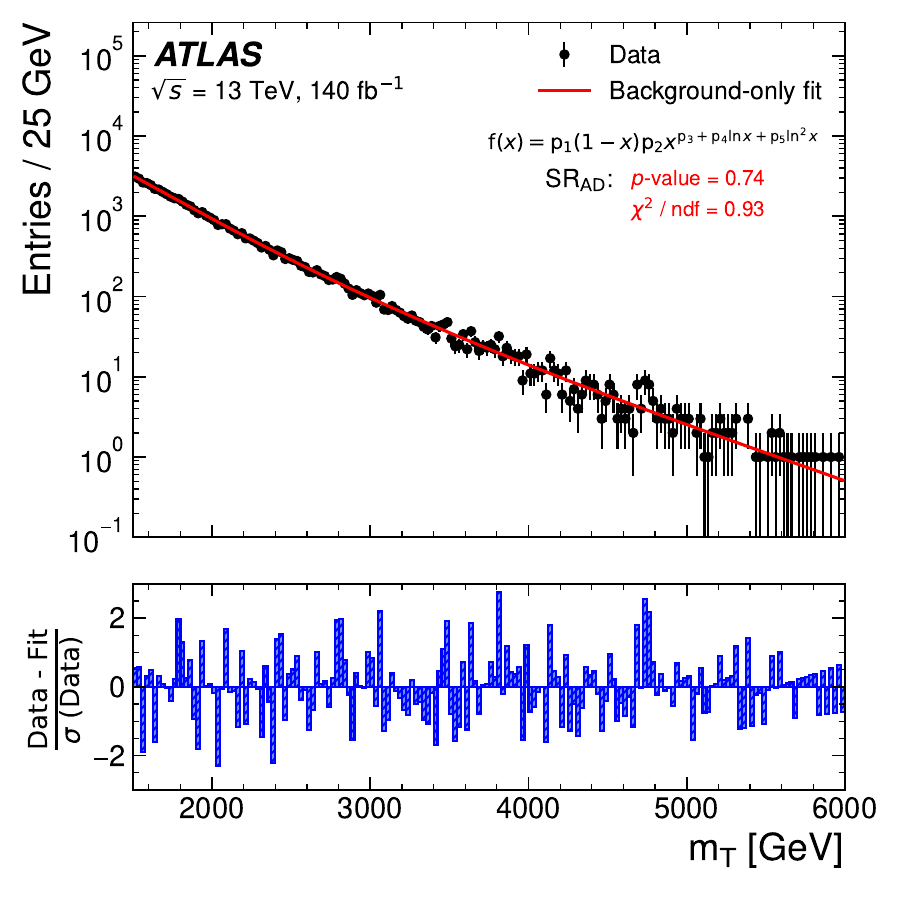}
\caption{}
\label{fig:bkg_fit_ANTELOPE}
\end{subfigure}
\caption{Distribution of data in the \mt~spectrum of the (a) PFN and (b) ANTELOPE signal regions. No significant excesses are found and good fit quality is obtained.}
\label{fig:unblinded_mt}
\end{figure}

Upper limits on the signal production cross section $\sigma$ times branching ratio $B$, are set at 95\% confidence level (CL) using the CL$_{s}$ method~\cite{Read:2002hq}.
Figure~\ref{fig:limits_1D} shows the 95\% CL observed and expected limits obtained from fitting the unblinded SR.
As the data were found to be compatible with background prediction, a good agreement of observed and expected limits is seen.
The analysis delivers the best limits for low \rinv~signal points, with a loss of exclusion power at higher \rinv~points due to the very broad signal shape in \mt.
Values of \mz~from 2000 to 3200 GeV are excluded to varying \rinv~fractions ranging from 0.37 to 0.2, respectively.
Figure~\ref{fig:limits_2D} shows the expected and observed exclusion to 95\% confidence in the two-dimensional plane of \mz~and \rinv.
The analysis has the strongest exclusion power for low values of \rinv, where the signature is more visible, and lower masses, where the production cross section of the signal is higher.
These exclusions are the first set by ATLAS on $Z'$-mediated semi-visible jet signals.

\begin{figure}[!t]
\centering
\begin{subfigure}[b]{0.45\textwidth}
\centering
\includegraphics[width=\textwidth]{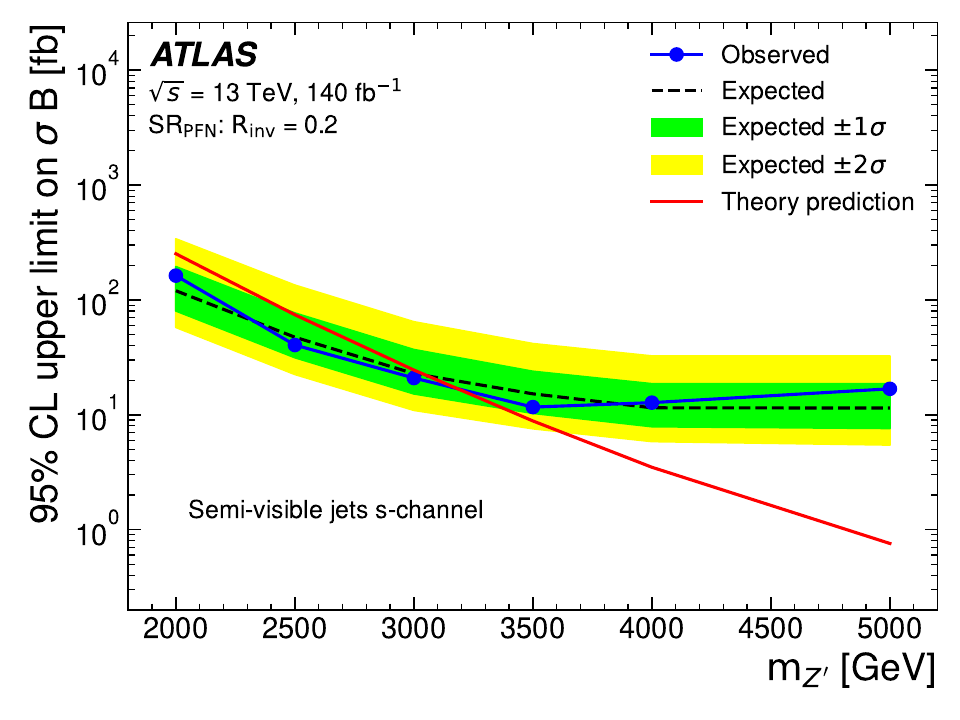}
\caption{}
\label{fig:limit_rinv_0p2}
\end{subfigure}
\hfill
\begin{subfigure}[b]{0.45\textwidth}
\centering
\includegraphics[width=\textwidth]{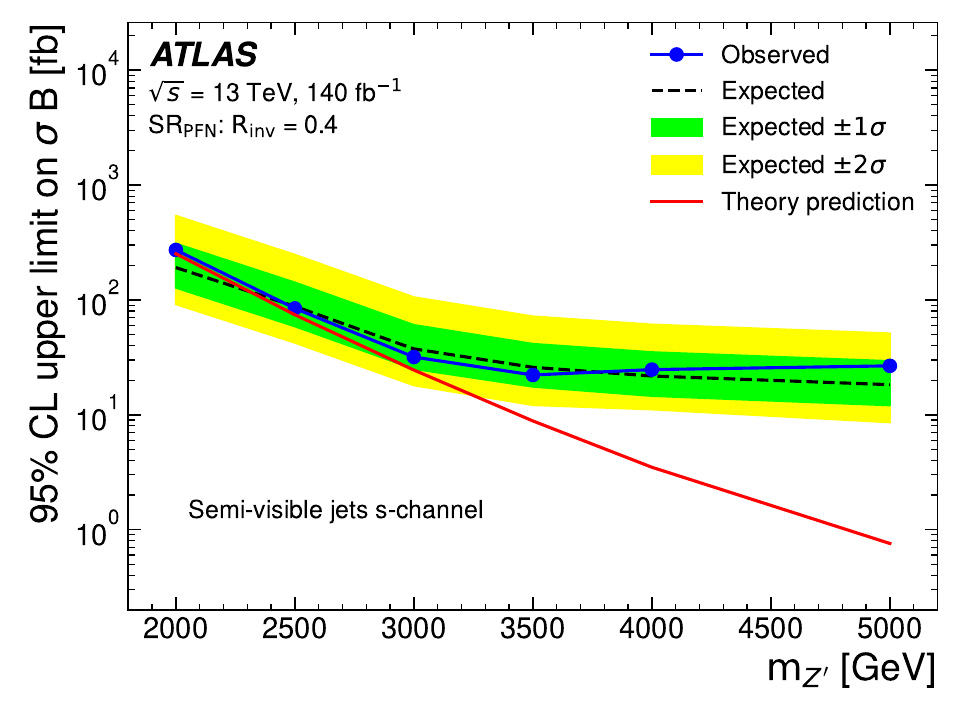}
\caption{}
\label{fig:limit_rinv_0p4}
\end{subfigure}

\vspace{0.5cm} %

\begin{subfigure}[b]{0.45\textwidth}
\centering
\includegraphics[width=\textwidth]{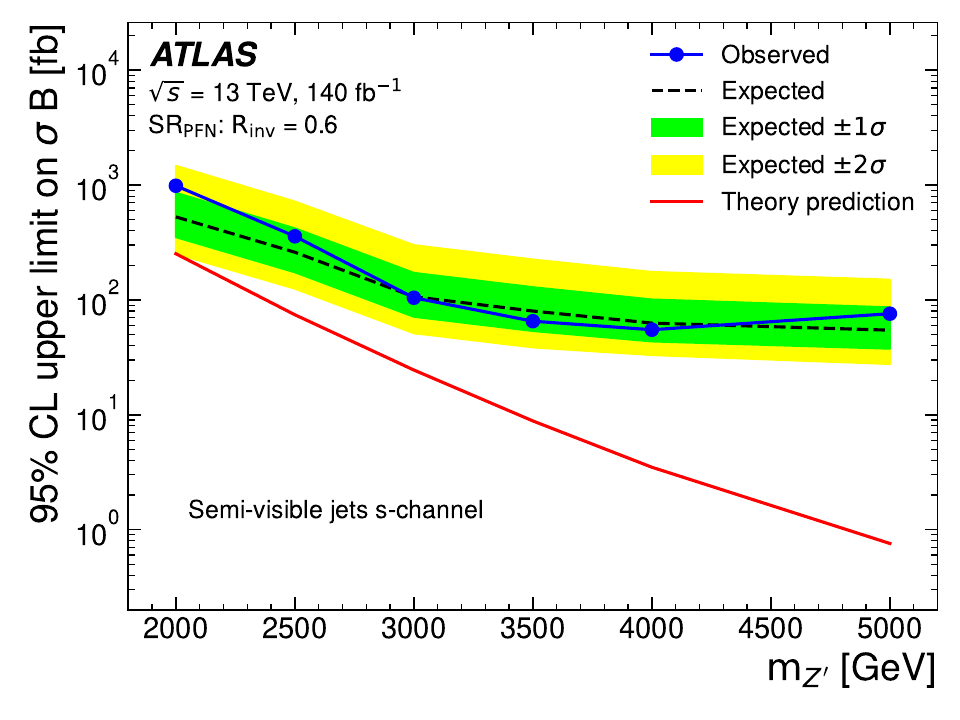}
\caption{}
\label{fig:limit_rinv_0p6}
\end{subfigure}
\caption{Observed (solid) and expected (dashed) limits at 95\% CL on the production cross section $\sigma$ times branching ratio $B$ of the $Z'$ mediator as a function of $Z'$ mass in GeV, for simulated \rinv~of (a) 0.2, (b) 0.4, and (c) 0.6. The dark (light) band around the limits represents the $\pm$1 (2) uncertainty band. The predicted theoretical cross section is also included in red obtained from the model discussed in Section~\ref{sec:datamc}.
\label{fig:limits_1D}}
\end{figure}

\begin{figure}[!t]
\centering
\includegraphics[width=0.65\textwidth]{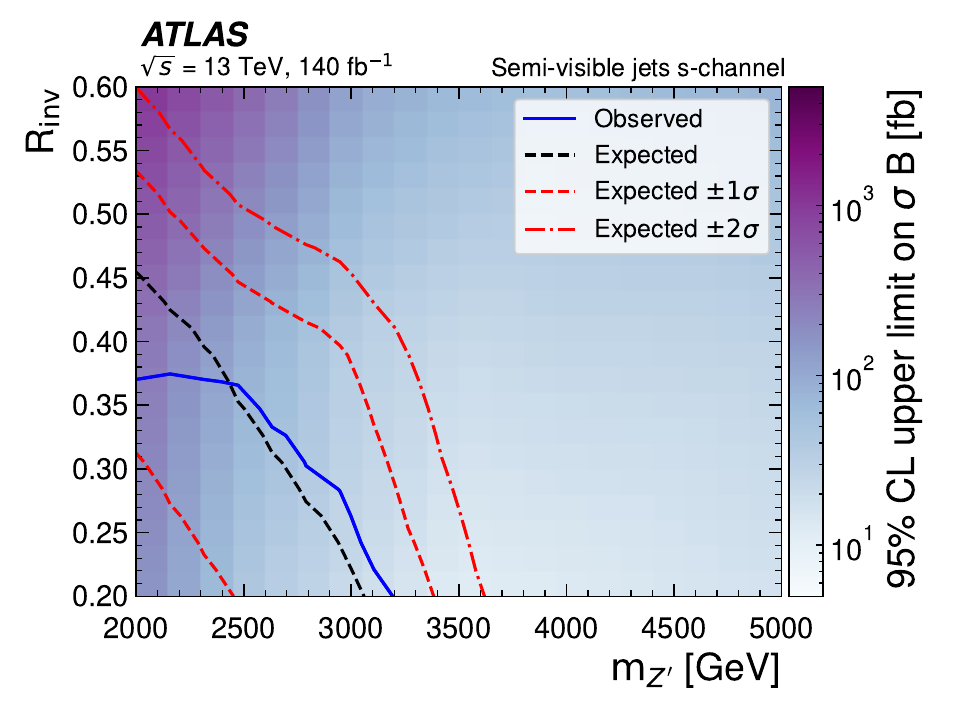}
\caption{The expected and observed exclusion contours at 95\% CL for the generated signal parameterized by \rinv~on the vertical axis and \mz~on the horizontal axis. The dashed line indicates the expected limit and the solid line corresponds to the observed limit derived from the unblinded \srpfn.
\label{fig:limits_2D}}
\end{figure}

While the ANTELOPE region is not used to set limits on the signal grid, it can still be used to compare the reach of the PFN SR to a less model-dependent approach.
This is done through signal injection into the ANTELOPE data to determine the 95\% upper limit on the signal strength $\mu$ of both the SVJ and alternate signal models discussed in Section~\ref{sec:datamc}.
The background is taken to be the background-only functional fit.
The 95\% upper limit is defined as the amount of signal at which BumpHunter returns a 2$\sigma$ significance.
As this check uses only BumpHunter, none of the signal systematics are applied, meaning these limits cannot be compared to dedicated analyses for theses signals.
Figure~\ref{fig:antelope_xsecLimit} shows the resulting limit plot comparing the PFN and ANTELOPE SRs.
As expected, the dedicated PFN region designed for the SVJ signals provides the better limit on the SVJ grid points by approximately a factor of two.
However, the ANTELOPE region takes over with nearly an order of magnitude improvement on the emerging jet and gluino R-hadron alternate signals.
This demonstrates the generality of the ANTELOPE tool and its utility to provide anomaly detection capability in this search.

\begin{figure}[!t]
\centering
\includegraphics[width=0.85\textwidth]{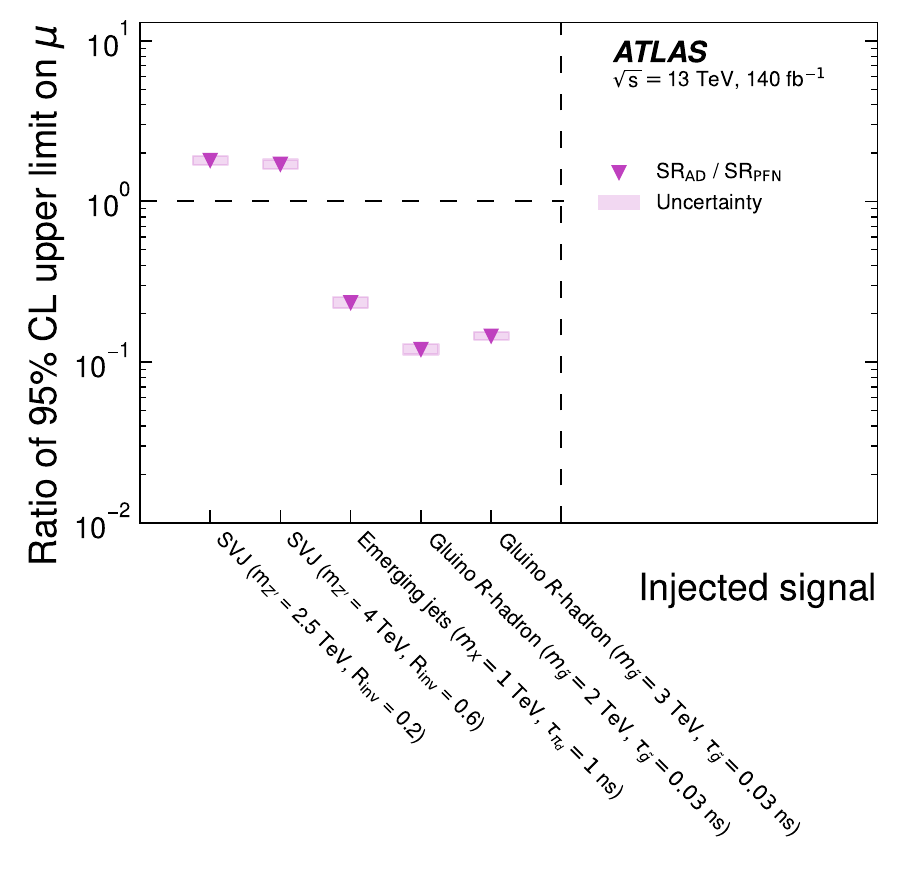}
\caption{The ratio of the 95\% upper limit on the signal strength for five benchmark signal processes, including two SVJ signals and three alternate models, comparing the PFN and ANTELOPE signal regions. The signal strength is obtained by injecting signal into the observed data until 2$\sigma$ sensitivity is achieved as determined by the \textsc{BumpHunter} interval $p$-value. The background is taken to be the result of the background-only functional fit on each unblinded region. The uncertainty band reflects the standard deviation of obtained signal strength values for 100 trials. SVJ signals are labeled by their \mz~and \rinv~values, gluino signals by the gluino mass and lifetime, and the emerging jet signal has $m_{Z'}$ = 1000 GeV and $c\tau_{\pi_d}$ = 1 mm. The PFN SR provides the best limit for the SVJ signal models which it saw in training, while the ANTELOPE region takes over for alternate signal models, indicating its lower degree of model dependence.
\label{fig:antelope_xsecLimit}}
\end{figure}

\clearpage

% End of text imported from the .//sections/results.tex input file

% The next lines are included from the .//sections/conclusions.tex input file
\section{Conclusions}
\label{sec:conclusions}

A search has been presented for strongly coupled dark sector particles in semi-visible jet signatures.
The SVJ signal is generated through the resonant production of a heavy $Z'$ mediator decaying to two dark quarks which decay to jets with a fraction \rinv~of constituent particles that are invisible to the detector.
Signal-like events are selected with the assistance of two machine learning tools which focus on track-level permutation invariant modeling of the events.
One of these, a ParticleFlow Network, provides the highest sensitivity to the simulated SVJ signal, while the second tool ANTELOPE focuses on anomaly detection through generalized sensitivity to a wide variety of signal models.
The analysis searches for excesses on top of a smoothly falling \mt~spectrum, which is estimated with a five-parameter functional fit.
No significant excesses are observed in the unblinded PFN and ANTELOPE regions, and upper limits are set on the production cross section of the $Z'$.
Signal points with \mz~between 2000 and 3200 GeV are excluded at 95\% CL for \rinv~ranging from 0.2 to 0.37.

% End of text imported from the .//sections/conclusions.tex input file

%
%
%
%
%
\section*{Acknowledgements}
%

%

% The next lines are included from the .//acknowledgements/Acknowledgements.tex input file
%
%

%
%

We thank CERN for the very successful operation of the LHC and its injectors, as well as the support staff at
CERN and at our institutions worldwide without whom ATLAS could not be operated efficiently.

The crucial computing support from all WLCG partners is acknowledged gratefully, in particular from CERN, the ATLAS Tier-1 facilities at TRIUMF/SFU (Canada), NDGF (Denmark, Norway, Sweden), CC-IN2P3 (France), KIT/GridKA (Germany), INFN-CNAF (Italy), NL-T1 (Netherlands), PIC (Spain), RAL (UK) and BNL (USA), the Tier-2 facilities worldwide and large non-WLCG resource providers. Major contributors of computing resources are listed in Ref.~\cite{ATL-SOFT-PUB-2025-001}.

We gratefully acknowledge the support of ANPCyT, Argentina; YerPhI, Armenia; ARC, Australia; BMWFW and FWF, Austria; ANAS, Azerbaijan; CNPq and FAPESP, Brazil; NSERC, NRC and CFI, Canada; CERN; ANID, Chile; CAS, MOST and NSFC, China; Minciencias, Colombia; MEYS CR, Czech Republic; DNRF and DNSRC, Denmark; IN2P3-CNRS and CEA-DRF/IRFU, France; SRNSFG, Georgia; BMBF, HGF and MPG, Germany; GSRI, Greece; RGC and Hong Kong SAR, China; ICHEP and Academy of Sciences and Humanities, Israel; INFN, Italy; MEXT and JSPS, Japan; CNRST, Morocco; NWO, Netherlands; RCN, Norway; MNiSW, Poland; FCT, Portugal; MNE/IFA, Romania; MSTDI, Serbia; MSSR, Slovakia; ARIS and MVZI, Slovenia; DSI/NRF, South Africa; MICIU/AEI, Spain; SRC and Wallenberg Foundation, Sweden; SERI, SNSF and Cantons of Bern and Geneva, Switzerland; NSTC, Taipei; TENMAK, T\"urkiye; STFC/UKRI, United Kingdom; DOE and NSF, United States of America.

Individual groups and members have received support from BCKDF, CANARIE, CRC and DRAC, Canada; CERN-CZ, FORTE and PRIMUS, Czech Republic; COST, ERC, ERDF, Horizon 2020, ICSC-NextGenerationEU and Marie Sk{\l}odowska-Curie Actions, European Union; Investissements d'Avenir Labex, Investissements d'Avenir Idex and ANR, France; DFG and AvH Foundation, Germany; Herakleitos, Thales and Aristeia programmes co-financed by EU-ESF and the Greek NSRF, Greece; BSF-NSF and MINERVA, Israel; NCN and NAWA, Poland; La Caixa Banking Foundation, CERCA Programme Generalitat de Catalunya and PROMETEO and GenT Programmes Generalitat Valenciana, Spain; G\"{o}ran Gustafssons Stiftelse, Sweden; The Royal Society and Leverhulme Trust, United Kingdom.

In addition, individual members wish to acknowledge support from Armenia: Yerevan Physics Institute (FAPERJ); CERN: European Organization for Nuclear Research (CERN DOCT); Chile: Agencia Nacional de Investigaci\'on y Desarrollo (FONDECYT 1230812, FONDECYT 1240864); China: Chinese Ministry of Science and Technology (MOST-2023YFA1605700, MOST-2023YFA1609300), National Natural Science Foundation of China (NSFC - 12175119, NSFC 12275265); Czech Republic: Czech Science Foundation (GACR - 24-11373S), Ministry of Education Youth and Sports (ERC-CZ-LL2327, FORTE CZ.02.01.01/00/22\_008/0004632), PRIMUS Research Programme (PRIMUS/21/SCI/017); EU: H2020 European Research Council (ERC - 101002463); European Union: European Research Council (BARD No. 101116429, ERC - 948254, ERC 101089007), European Regional Development Fund (SMASH COFUND 101081355, SLO ERDF), Horizon 2020 Framework Programme (MUCCA - CHIST-ERA-19-XAI-00), European Union, Future Artificial Intelligence Research (FAIR-NextGenerationEU PE00000013), Italian Center for High Performance Computing, Big Data and Quantum Computing (ICSC, NextGenerationEU); France: Agence Nationale de la Recherche (ANR-21-CE31-0022, ANR-22-EDIR-0002); Germany: Baden-Württemberg Stiftung (BW Stiftung-Postdoc Eliteprogramme), Deutsche Forschungsgemeinschaft (DFG - 469666862, DFG - CR 312/5-2); China: Research Grants Council (GRF); Italy: Istituto Nazionale di Fisica Nucleare (ICSC, NextGenerationEU), Ministero dell'Università e della Ricerca (NextGenEU I53D23000820006 M4C2.1.1, SOE2024\_0000023); Japan: Japan Society for the Promotion of Science (JSPS KAKENHI JP22H01227, JSPS KAKENHI JP22H04944, JSPS KAKENHI JP22KK0227, JSPS KAKENHI JP23KK0245, JSPS KAKENHI JP24K23939); Norway: Research Council of Norway (RCN-314472); Poland: Ministry of Science and Higher Education (IDUB AGH, POB8, D4 no 9722), Polish National Science Centre (NCN 2021/42/E/ST2/00350, NCN OPUS 2023/51/B/ST2/02507, NCN OPUS nr 2022/47/B/ST2/03059, NCN UMO-2019/34/E/ST2/00393, UMO-2022/47/O/ST2/00148, UMO-2023/49/B/ST2/04085, UMO-2023/51/B/ST2/00920, UMO-2024/53/N/ST2/00869); Portugal: Foundation for Science and Technology (FCT); Spain: Ministry of Science and Innovation (MCIN \& NextGenEU PCI2022-135018-2, MICIN \& FEDER PID2021-125273NB, RYC2019-028510-I, RYC2020-030254-I, RYC2021-031273-I, RYC2022-038164-I); Sweden: Carl Trygger Foundation (Carl Trygger Foundation CTS 22:2312), Swedish Research Council (Swedish Research Council 2023-04654, VR 2021-03651, VR 2022-03845, VR 2022-04683, VR 2023-03403, VR 2024-05451), Knut and Alice Wallenberg Foundation (KAW 2018.0458, KAW 2022.0358, KAW 2023.0366); Switzerland: Swiss National Science Foundation (SNSF - PCEFP2\_194658); United Kingdom: Leverhulme Trust (Leverhulme Trust RPG-2020-004), Royal Society (NIF-R1-231091); United States of America: U.S. Department of Energy (ECA DE-AC02-76SF00515), Neubauer Family Foundation.

%
%

% End of text imported from the .//acknowledgements/Acknowledgements.tex input file

%
%
%
%
%
%

%
%
%
\printbibliography
\clearpage
% ATLAS Collaboration author list
% Reference date of EXOT-2021-19 is 2025-03-16
% Author list last updated on date 24-JUN-25
% Data extracted on 24-Jun-2025 for paper reference EXOT-2021-19
% at 9:19am
 
\begin{flushleft}
\hypersetup{urlcolor=black}
{\Large The ATLAS Collaboration}

\bigskip

\AtlasOrcid[0000-0002-6665-4934]{G.~Aad}$^\textrm{\scriptsize 104}$,
\AtlasOrcid[0000-0001-7616-1554]{E.~Aakvaag}$^\textrm{\scriptsize 17}$,
\AtlasOrcid[0000-0002-5888-2734]{B.~Abbott}$^\textrm{\scriptsize 123}$,
\AtlasOrcid[0000-0002-0287-5869]{S.~Abdelhameed}$^\textrm{\scriptsize 119a}$,
\AtlasOrcid[0000-0002-1002-1652]{K.~Abeling}$^\textrm{\scriptsize 55}$,
\AtlasOrcid[0000-0001-5763-2760]{N.J.~Abicht}$^\textrm{\scriptsize 49}$,
\AtlasOrcid[0000-0002-8496-9294]{S.H.~Abidi}$^\textrm{\scriptsize 30}$,
\AtlasOrcid[0009-0003-6578-220X]{M.~Aboelela}$^\textrm{\scriptsize 45}$,
\AtlasOrcid[0000-0002-9987-2292]{A.~Aboulhorma}$^\textrm{\scriptsize 36e}$,
\AtlasOrcid[0000-0001-5329-6640]{H.~Abramowicz}$^\textrm{\scriptsize 157}$
% ...
\bigskip
\\

% ...
$^{17}$Department for Physics and Technology, University of Bergen, Bergen; Norway.\\
% ...
$^{30}$Physics Department, Brookhaven National Laboratory, Upton NY; United States of America.\\
% ...
$^{36}$$^{(a)}$Facult\'e des Sciences Ain Chock, Universit\'e Hassan II de Casablanca;$^{(b)}$Facult\'{e} des Sciences, Universit\'{e} Ibn-Tofail, K\'{e}nitra;$^{(c)}$Facult\'e des Sciences Semlalia, Universit\'e Cadi Ayyad, LPHEA-Marrakech;$^{(d)}$LPMR, Facult\'e des Sciences, Universit\'e Mohamed Premier, Oujda;$^{(e)}$Facult\'e des sciences, Universit\'e Mohammed V, Rabat;$^{(f)}$Institute of Applied Physics, Mohammed VI Polytechnic University, Ben Guerir; Morocco.\\
% ...
$^{45}$Physics Department, Southern Methodist University, Dallas TX; United States of America.\\
% ...
$^{49}$Fakult\"{a}t Physik , Technische Universit{\"a}t Dortmund, Dortmund; Germany.\\
% ...
$^{55}$II. Physikalisches Institut, Georg-August-Universit\"{a}t G\"ottingen, G\"ottingen; Germany.\\
% ...
$^{104}$CPPM, Aix-Marseille Universit\'e, CNRS/IN2P3, Marseille; France.\\
% ...
$^{119}$$^{(a)}$New York University Abu Dhabi, Abu Dhabi;$^{(b)}$United Arab Emirates University, Al Ain; United Arab Emirates.\\
% ...
$^{123}$Homer L. Dodge Department of Physics and Astronomy, University of Oklahoma, Norman OK; United States of America.\\
% ...
$^{157}$Raymond and Beverly Sackler School of Physics and Astronomy, Tel Aviv University, Tel Aviv; Israel.\\
% ...
$^{a}$ Also at Affiliated with an institute formerly covered by a cooperation agreement with CERN.\\
$^{b}$ Also at An-Najah National University, Nablus; Palestine.\\
$^{c}$ Also at Borough of Manhattan Community College, City University of New York, New York NY; United States of America.\\
$^{d}$ Also at Center for Interdisciplinary Research and Innovation (CIRI-AUTH), Thessaloniki; Greece.\\
$^{e}$ Also at Centre of Physics of the Universities of Minho and Porto (CF-UM-UP); Portugal.\\
$^{f}$ Also at CERN, Geneva; Switzerland.\\
% ...

\end{flushleft}

% Created with Glance <Atlas.Glance@cern.ch>

%

%
%

%

%
%
%
%

\end{document}